\documentclass{aastex63}  
\usepackage[T1]{fontenc} 

\received{2020 Feb 24} 
\revised{2020 May 14}
\accepted{2020 May 22}
\submitjournal{AJ}

\shorttitle{Strengths of meteoroids}
\shortauthors{Borovi\v{c}ka et al.}

\begin{document}

\title{Two Strengths of Ordinary Chondritic Meteoroids as Derived from their Atmospheric Fragmentation Modeling}

\correspondingauthor{Ji\v{r}\'{\i} Borovi\v{c}ka}
\email{jiri.borovicka@asu.cas.cz}

\author[0000-0002-5569-8982]{Ji\v{r}\'{\i} Borovi\v{c}ka}
\affiliation{Astronomical Institute of the Czech Academy of Sciences,
Fri\v{c}ova 298, CZ-25165 Ond\v{r}ejov, Czech Republic}

\author{Pavel Spurn\'{y}}
\affiliation{Astronomical Institute of the Czech Academy of Sciences,
Fri\v{c}ova 298, CZ-25165 Ond\v{r}ejov, Czech Republic}

\author{Luk\'a\v{s} Shrben\'{y}}
\affiliation{Astronomical Institute of the Czech Academy of Sciences,
Fri\v{c}ova 298, CZ-25165 Ond\v{r}ejov, Czech Republic}

\begin{abstract}

The internal structure and strength of small asteroids and large meteoroids is poorly known.
Observation of bright fireballs in the Earth's atmosphere can prospect meteoroid structure by studying
meteoroid fragmentation during the flight. Earlier evaluations showed that meteoroid strength is
significantly lower than that of the recovered meteorites. We present detailed study of atmospheric
fragmentation of seven meteorite falls, all ordinary chondrites, and 14 other fireballs, where meteorite
fall was predicted but the meteorites, probably also ordinary chondrites, were not recovered.
All observations were made by the autonomous observatories of the European Fireball Network and 
include detailed radiometric light curves. A model, called the semi-empirical fragmentation model, was developed to fit
the light curves and decelerations. Videos showing individual fragments were available in some cases.
The results demonstrated that meteoroids do not fragment randomly but in two distinct phases.
The first phase typically corresponds to low strengths of 0.04 -- 0.12 MPa. In 2/3 of cases,
the first phase was catastrophic or nearly catastrophic with at least 40\% of mass lost.
The second phase corresponds to 0.9 -- 5 MPa for confirmed meteorite falls and to somewhat lower strengths,
from about 0.5 MPa for smaller meteoroids. All these strengths are lower than tensile
strengths of ordinary chondritic meteorites cited in the literature, 20 -- 40 MPa.
We interpret the second phase as being due by cracks in meteoroids and the first phase
as separation of weakly cemented fragments, which reaccumulated at surfaces of asteroids after 
asteroid collisions.

\end{abstract}

\keywords{meteoroids --- meteors ---asteroids --- meteorites}

\section{Introduction} 
\label{intro}

There are two main methods of studying interplanetary material in the vicinity of the Earth: astronomical and geochemical.
The astronomical method relies primarily on telescopic observations of asteroids and comets, both from ground and space. 
In favorable cases, optical observations can be supplemented by radar investigation. The obtained data include orbits, rotational properties, 
shapes, albedos, sizes, and reflectance spectra of asteroids. For active comets, their emission spectra and properties 
of the released gases and dust can be obtained. The geochemical method is applicable only to meteorites and, in few recent
cases, to samples returned by spacecraft. The advantage is that mineralogical, chemical, 
and physical properties of meteorites and returned samples can be studied in laboratory in detail.

Of particular interest are physical properties and internal structure of asteroids. Their knowledge is important to understand
evolutionary history of asteroids, and inner Solar System in general, as well as to
evaluate the consequences of potential asteroid collision with the Earth. Physical properties are also to be
considered in any attempt of deflecting an asteroid from impact trajectory or in an effort for asteroid mining.
Telescopic observations, however, provide limited information about internal structure of asteroids.
The existence of a spin barrier at the rotation period of 2.4 hours was revealed for asteroids in the size range from
$\sim 200$ meters to $\sim 10$ km \citep{Pravec, Hestroffer}, suggesting that most asteroids in this size range may be aggregates
of smaller blocks held together only by their mutual gravity -- so called rubble piles. Asteroids smaller than 200 meters
have wide range of rotational periods and their internal strength is not restricted by these observations.

The most common type of meteorites, ordinary chondrites, which represent 81\% of all meteorite falls \citep{chapter2}, 
are hard objects. Individual measured strengths vary widely, nevertheless, typical compressive strengths are between 100 -- 200 MPa 
and tensile strengths between 20 -- 40 MPa, i.e.\
comparable to the strongest terrestrial rocks \citep{Slyuta, Flynn, Ostrowski}. 
Meteorites are delivered by meteoroids and small asteroids in the size range from decimeters to decameters 
(thereafter collectively called 'meteoroids')
and represent their strongest parts, which survived the atmospheric passage. 
It is well known that meteoroids are subject to fragmentation during their interaction with the atmosphere.
The sole fact that vast majority of meteorite falls produce more than one meteorite, sometimes of the order of
thousand pieces \citep[e.g.][]{Mbale,BuzzardC}, is a sufficient proof of meteoroid atmospheric fragmentation. 
Incoming meteoroids are therefore weaker than the recovered meteorites.

In this work we use atmospheric fragmentation to evaluate strengths of meteoroids. This idea is not new. 
\citet{Fadeenko}, among the first, considered that meteoroid will fragment when the dynamic pressure acting
on the leading surface,
\begin{equation}
p = \rho v^2,
\label{dynpres}
\end{equation}
where $\rho$ is atmospheric density and $v$ is meteoroid velocity, exceeds the strength of the material. 
Other authors used the same relation when either modeling the meteoroid atmospheric entry 
\citep[e.g.][]{Baldwin, HillsGoda, Svetsov, Bland, Register} or
inferring meteoroid strengths from observations \citep[e.g.][]{Trigo,Popova}. Note that the meteoroid strength is not precisely
defined in this approach. Various types of material strength were discussed by \citet{Holsapple}.
The fragmentation strength derived from the dynamic pressure is usually considered
to correspond to the tensile strength of the material \citep[e.g.][]{Trigo}, but a recent calculation showed 
that the shear strength may be the most relevant, at least for large bodies \citep{Robertson}. 

\citet{Ceplecha} studied fragmentation of 51 fireballs photographed by the US Prairie Network. 
Only fireball dynamic data, i.e.\ lengths along the trajectory as a function of time were used. Fragmentations
were therefore found 
on the basis of increased deceleration after a sudden mass loss. The method was able to reveal one dominant 
fragmentation point in 19 cases. Eleven fireballs produced
so-called inverse solutions, which were interpreted as multi-fragmentation events. The positions of the
fragmentations could not be found in these cases. 
The strengths for single-fragmentation events were found in the range 0.05 -- 1.2 MPa. 
One exceptional meteoroid survived without significant fragmentation up to 5 MPa.

\citet{Popova} evaluated fragmentation in 11 instrumentally observed meteorite falls. Fragmentation data were 
compiled from original sources. The used dataset was therefore very heterogeneous with fireball data of varying
quality obtained by various techniques. Various signatures were used to reveal fragmentation points along the
trajectory. The resulting message was, nevertheless, clear. Incoming meteoroids and small asteroids have very
low strengths in comparison with meteorites.

In this paper we compile analyses of 7 instrumentally observed falls of ordinary chondrites 
\citep[one is the same as in][six are new]{Popova}. 
In all cases the fireball data are sufficient for detailed fragmentation modeling. In
addition to known trajectories and velocities, radiometric light curves with high temporal resolution and high
dynamic range are available.  In some cases, dynamic data, i.e.\ fireball decelerations, are also available along the whole
trajectories. In addition, we
analyze 14 other fireballs, which certainly also dropped meteorites of masses of at least several tens of grams
but the meteorites were, unfortunately not recovered. In these cases, both radiometric curves and
good deceleration data are available.  

The instrumentation and data are described in Section~\ref{instrum}. 
The fragmentation model used to reveal meteoroid fragmentation behavior from fireball data is presented
in Section~\ref{model}. The modeling results are given in Section~\ref{results}, separately for the
confirmed meteorite falls and for the additional fireballs. For the latter, orbital data and coordinates of the
strewn fields are also provided. For the confirmed falls, these data can be found in the original papers.
The revealed fragmentation behavior and its implication for the structure of ordinary chondritic meteoroids 
is discussed in Section~\ref{discussion} and summarized in Section~\ref{summary}.

\newpage

\section{Instrumentation and data}
\label{instrum}

The data analyzed here were primarily obtained in the scope of the European Fireball Network (EN), a long-term project
of observing fireballs over central Europe \citep{Spurny_EN}, in 2009--2018. During that period, the main instrument
of the network changed from the older Autonomous Fireball Observatories \citep[AFO, described in][]{Spurny_EN} to 
the modern Digital Autonomous Fireball Observatories \citep[DAFO, described briefly in][]{Spurny_Taurids}. 
The purpose of both types of instruments was to take images of the whole sky during each night, when the sky was at least
partly clear. The main difference is that AFO used photographic film and took usually one exposure per night while
DAFO uses two digital DSLR cameras Canon 6D and is taking multiple exposures 35 s long. To enable measurement
of fireball velocities, AFO employed a mechanical rotating shutter placed in front of the film. DAFO uses an LCD shutter
placed behind the lens. The shutter break frequency is similar in both cases, 15 Hz for AFO and 16 Hz for DAFO.

Besides avoiding the laborious manipulation with photographic films, the advantages of DAFO are higher sensitivity and
better performance in difficult conditions (moonlit nights, partly cloudy nights, and twilight periods). The point-like
appearance of stars and higher number of stars in DAFO images (especially in regions close to horizon) makes the astrometric
reduction easier and more reliable. The disadvantage of DAFO is lower dynamic range, which makes the photometry of 
bright fireballs difficult. Nevertheless, the main photometric instrument is radiometer, which is part of both AFO and DAFO.

The radiometer is a photomultiplier tube with flat entry aperture without any optics, directed to zenith. It takes the measurement
of the total brightness of the sky 5000 times per second. The dynamic range is 20 bits, providing information 
about the luminosity of fireballs in linear scale for fireballs of apparent magnitudes from about $-2$ to about $-17$. 
Fireballs brighter than $-17$ mag (i.e.\ superbolides) can be reliably measured from more distant stations, where their 
apparent magnitude is lower. The radiometer provides intensity in relative units and the zero point must be determined for each
fireball using photographic data (which provide absolute fireball photometry by comparison with stars). 
For this purpose, the non-saturated part of the
photographic data is used (usually the part of the light curve when the fireball magnitude was between $-5$ and $-8$).
The response of the radiometer as a function of zenith angle was measured in laboratory. To join the radiometric and 
photographic data easily together, a time mark is produced by the DAFO LCD shutter every second by skipping one 
interruption (i.e.\ making one dash on the fireball image three times longer). Both the LCD and radiometer time are controlled
by the GPS signal to keep the absolute time correct with sub-millisecond precision. In case of AFO, visible features 
on the photographic light curve had to be compared with the radiometric curve to determine absolute timing.

The DAFOs started to be deployed at the stations of the network at the end of 2013. By the end of 2014, all stations
in the Czech Republic were equipped with DAFOs. DAFOs and AFOs were then run in parallel for the next few years 
(depending on station). In addition, 
several new DAFO stations were build in 2015--2018. By the end of 2018, only DAFOs were used at all 14 EN stations in the
Czech Republic, 3 in Slovakia, and 1 in Austria. Old type mirror cameras \citep{Oberst} were used in Germany as 
part of the EN.

\begin{table*}
\caption{Meteorite falls analyzed in this study}
\label{meteoritelist}
\begin{flushleft}
\vspace{-2em}
\begin{tabular}{llllllllll}
\hline \hline
Name & Type & Date & Latitude & Longitude & Country & Entry mass & Recovered & No. of & Ref.\\[-0.75ex] 
 & & & & & & from model & mass & frag- \\
 & & & \degr N & \degr E & & kg & kg& ments\\
\hline
Jesenice & L6 & 2009 Apr 9 &  46.421 & 14.052 & Slovenia & 250 & 3.61 & 3 & [1,2]\\
Ko\v{s}ice & H5 & 2010 Feb 28 & 48.757 & 21.160 & Slovakia & 4000 & 11.3 &  218 & [3,4] \\
Kri\v{z}evci & H6 & 2011 Feb 4 & 46.039 & 16.590 & Croatia & 53 & 0.29 & 1 & [5] \\
\v{Z}\v{d}\'{a}r nad S\'{a}zavou & L3.9 & 2014 Dec 9 & 49.508 & 15.963 & Czech R. & 150 & 0.087 & 3 & [6,7] \\
Stubenberg & LL6 & 2016 Mar 6 & 48.306 & 13.093 & Germany & 450 & 1.47 & 6 & [8,9,P] \\
Hradec Kr\'{a}lov\'{e} & LL5 & 2016 May 17 & 50.301 &  15.728 & Czech R. & 90 & 0.13 & 1 & [P] \\
Renchen &  L5-6 & 2018 Jul 10 & 48.610 & \ \,7.948 & Germany  & 17 & 1.23 & 6 & [10,P] \\
\hline
\end{tabular}
\end{flushleft}
\vspace{-1em}
\tablecomments{References: [1]~\citet{Jesenice}, [2]~\citet{Bischoff_Jesenice}, [3]~\citet{Kosice}, [4]~\citet{Toth_Kosice}, [5]~\citet{Krizevci}
[6]~\citet{Zdar}, [7]~\citet{Zdar_Kalasova}, [8]~\citet{Spurny_abstract}, [9]~\citet{Bischoff_Stubenberg},
[10]~\citet{Bischoff_Renchen} [P]~Papers in preparation}
\end{table*}

\subsection{Observed meteorite falls}

One of the purposes of the European Fireball Network is to observe meteorite dropping fireballs. 
The data are used to predict
the location of meteorites, to study the interaction of the meteoroid with the atmosphere, and to compute
the pre-encounter heliocentric orbit. Recovering the meteorites
is, however, not easy. While the location of the fireball luminous trajectory can be determined with the
precision of the order of tens of meters, the dark flight (starting usually from heights between 20--30 km)
cannot be observed and the final meteorite location depends on exact meteorite mass and shape as well 
as on upper atmosphere winds. Taking winds from meteorological models and assuming spherical
shape of the meteorite, the location can be computed as a function of meteorite mass (more exactly,
a parameter combining mass and density). In case of good dynamic data toward the end of the luminous
trajectory, meteorite mass can be estimated from the observed speed and deceleration. Fragmentation 
model (Section~\ref{model}) can provide likely mass range for additional smaller meteorites and the whole
strewn field can be modeled. Depending on the geometry (especially the slope of the trajectory), meteoroid
fragmentation heights, wind speeds and directions, and meteorite mass range, strewn fields can have 
various sizes and shapes. Typically, the area of the highest probability to be searched is a strip of a couple
of hundreds of meters wide and several kilometers up to tens of kilometers long. 
The uncertainty in the position of the largest fragment is
typically a few hundreds of meters. Finding a single stone in the European landscape is
a challenge. The experience shows that the more favorable cases are those producing large number of small meteorites.
There is a chance to find at least some of them.

EN cameras have so far obtained data for 12  recovered meteorites (from about 30 meteorites with known trajectory worldwide)
and 10 of them are ordinary chondrites. Radiometric light curves are available for 8 of them, but in case 
of the Ejby meteorite fall in Denmark \citep{Ejby},  the quality of the light curve is not sufficient for fragmentation modeling
due to large distance of the camera from the fireball. 
Seven meteorite falls were therefore modeled. They are listed in Table~\ref{meteoritelist}. Meteorite name,
classification, date of fall, coordinates of the largest recovered fragment, entry mass estimated from the 
fragmentation modeling (Section~\ref{results_meteorites}), total recovered mass, and the number of recovered fragments
are given.   

The Jesenice and Kri\v{z}evci meteorite falls occurred also relatively far from the EN cameras but radiometric curves are good.
EN cameras were combined in these cases
with Slovenian and Croatian cameras for as complete description of the fireballs  as possible \citep{Jesenice, Krizevci}. 
For Jesenice, there are no deceleration data but the trajectory, entry speed, and light curve could be determined well. 
The first Jesenice meteorite was found casually before the exact fireball trajectory was computed. The Kri\v{z}evci meteorite
was found by a Croatian group on the basis of a preliminary trajectory computed from Croatian cameras.

The Ko\v{s}ice meteorite fall occurred in bad weather when all EN cameras were
clouded out. The trajectory, velocity, and deceleration were determined from three casual video records extracted from security
cameras in Hungary \citep{Kosice}. Of course, the precision is lower than from dedicated cameras. 
But one video shows also a fragment following the main body towards the end. 
Thanks to the extreme brightness of this superbolide, good radiometric curves were 
obtained by EN cameras through thick clouds (in full Moon night!).

\begin{table}
\caption{Basic parameters of trajectories of meteorite falls}
\label{meteoritetraj}
\begin{tabular}{lllllr}
\hline \hline
Name & Begin.  & End & Zenith & Entry & Maximum \\[-0.75ex] 
         & Height  & Height  & angle & speed & magnitude \\
         & km    & km       & \degr & km/s &\\ \hline
Jesenice & 88. & $\sim$18.\tablenotemark{a} & 31. & 13.78 & $-15.~~~$\\ 
Ko\v{s}ice & N/O\tablenotemark{b} & 17.4 &  30.2 & 15.0 &  $-18.~~~$ \\
Kri\v{z}evci & 98.10 & 21.81 & 24.6 & 18.21 & $-13.7~~$ \\
\v{Z}\v{d}\'{a}r n.\ S. & 98.06 & 24.71 & 64.8 & 21.89 & $-15.3~~$ \\
Stubenberg & 85.92 & 17.19 & 19.6 & 13.91 & $-15.4~~$ \\ 
Hradec K. & 74.34 & 23.54 & 42.5 & 13.31 & $-11.5~~$ \\
Renchen & 80.40 & 18.47 & 11.9 & 18.62 & $-13.4~~$ \\
\hline
\end{tabular}
\tablenotetext{a}{Fireball end not directly observed. End height estimated from the light curve.} \vspace{-1ex}
\tablenotetext{b}{Fireball beginning not observed. Fireball entered the field of view at a height of 68.3 km.}
\end{table}

The last four meteorites were recovered on the basis of observations by EN cameras. \v{Z}\v{d}\'{a}r nad S\'{a}zavou
fell in the middle of the network and has excellent data. Stubenberg and Hradec Kr\'{a}lov\'{e} were affected by bad
weather. On some stations, only parts of the fireballs were recorded between clouds. The same is valid for Renchen,
where, moreover, the dynamics was measurable only on a German camera with lower resolution. In all cases,
nevertheless, the trajectory, orbit, and light curve were determined reliably. The only missing data are decelerations
at the end of the trajectories. 
Basic parameters of the trajectories are given in Table~\ref{meteoritetraj}.

\begin{table*}
\caption{Trajectories of fireballs analyzed in this study}
\label{fireballlist}
\begin{flushleft}
\vspace{-2em}
\begin{tabular}{llllllllllr}
\hline \hline
Name & Time & \multicolumn{3}{c}{Beginning point}  & \multicolumn{3}{c}{End point} & Zenith & Entry & Maximum \\[-0.75ex] 
(Date) &UT & Longitude &  Latitude  & Height &Longitude &  Latitude  & Height           & angle & speed & magnitude \\
 &hms & \degr E & \degr N & km &  \degr E & \degr N & km & \degr & km/s &\\ \hline
2015\,Jun\,02&215119 &12.0153 &49.8729 &85.09 &11.9737 &50.0746 &24.30 &20.4 &16.28 &$-9.7~~$   \\
2015\,Aug\,26&233145 &11.8268 &48.3390 &90.04 &12.3564 &48.9995 &28.43 &53.3 &19.36 &$-11.5~~$\\
2016\,Nov\,10&022429 &20.6512 &48.6528 &91.86 &21.0221 &48.3100 &20.37 &33.3 &23.49 &$-12.5~~$\\
2016\,Dec\,07&041110 &13.5418 &49.7472 &86.70 &15.0116 &49.7543 &28.64 &61.3 &21.03 &$-10.3~~$\\
2017\,Feb\,24&190640 &14.0102 &48.3757 &84.22 &13.3198 &48.5257 &31.38 &45.6 &17.82 &$-9.6~~$\\
2017\,Feb\,27&023122 &13.4461 &49.1474 &95.44 &14.2898 &49.6462 &27.95 &50.9 &31.24 &$-12.4~~$\\
2017\,Nov\,14&164658 &11.0091 &50.0767 &94.71 &07.4990 &50.2612 &26.17 &74.3 &19.77 &$-12.8~~$\\
2017\,Dec\,17&171520 &14.6954 &49.2097 &77.89 &14.4618 &49.0219 &21.18 &25.4 &13.32 &$-9.4~~$\\
2018\,Jan\,18&182623 &14.6084 &49.8803 &87.62 &14.3817 &49.3879 &26.39 &43.2 &20.06 &$-10.2~~$\\
2018\,Apr\,08&184736 &17.5960 &46.9418 &88.65 &16.6598 &46.1782 &25.05 &59.6 &16.45 &$-12.7~~$\\
2018\,May\,23&194647 &17.0983 &49.9734 &80.37 &17.1427 &49.5864 &23.47 &37.3 &12.92 &$-9.4~~$\\
2018\,Sep\,11&214648 &15.6510 &47.0533 &91.43 &15.6134 &47.3531 &25.92 &27.2 &23.65 &$-14.0~~$\\
2018\,Oct\,08&195513 &14.4817 &50.0272 &82.05 &14.2281 &50.3284 &23.51 &33.3 &13.98 &$-9.1~~$\\
2018\,Nov\,29&041019 &16.5184 &46.6258 &90.73 &16.5928 &45.9443 &22.45 &48.1 &25.82 &$-12.5~~$\\
\hline
\end{tabular}
\vspace{-1em}
\end{flushleft}
\end{table*}

\subsection{Additional fireballs}

To enlarge the statistics, the analysis of 7 meteorite falls was supplemented by the analysis of 14 other fireballs,
which almost certainly also dropped meteorites (at least small ones with masses of the order of tens of grams)
but the meteorites were not recovered. All these fireballs were observed very well, including dynamic data along the
whole trajectories. All of them are of type I according to the classification of \citet{CepMac76} and were therefore
produced by stony meteoroids, most probably ordinary chondrites. We cannot exclude presence of enstatite chondrites
or some strong achondrites but 
the presence of
 carbonaceous chondrites, which typically belong to type II, is unlikely, and irons, which 
belong to type III \citep{Vojacek_iron}, can be excluded. 
On statistical basis, 13 of the 14 type I fireballs should be ordinary chondrites,
since from ordinary and enstatite chondrites and achondrites, ordinary chondrites form 93\% of meteorite falls. 

The list of the fireballs is given in Table~\ref{fireballlist}. 
The fireballs will be referred to according to their date of appearance. The listed time is valid for the fireball beginning.
The official code of the fireball can be constructed from the date and time in the form ENddmmyy\_hhmmss,
e.g.\ EN020615\_215119 for the first one.
Geographical coordinates of the observed beginning and end points take into account Earth curvature and the estimated curvature
of the trajectory due to gravity. The zenith angle, i.e. zenith distance of the radiant 
($0\degr$ means vertical trajectory; $90\degr$ means horizontal flight), is valid for an average
point along the trajectory. Finally, the speed at the top of the atmosphere and the maximum absolute (i.e.\ at the
distance of 100 km) stellar magnitude are given.

The fireball data are mainly based on the observations by DAFOs. In seven cases, supplementary
video cameras installed at two EN stations (Ond\v{r}ejov and Kun\v{z}ak) in 2016--2018 were also used. 
These are Dahua surveillance cameras and provide 20 frames per second 
in resolution $2688\times1520$ pixels with field of view of $56\degr\times32\degr$. In the final
configuration, 13 cameras are continuously working at each station. Video cameras provided additional valuable
dynamic data, especially for fireball 2017\,Nov\,14, where the dynamics was fully based on video data because of too slow angular motion
at all DAFO stations, which made the shutter breaks hardly measurable. In other three cases (2017\,Dec\,17, 2018\,Apr\,08, 2018\,May\,23),
video data contained fragments following the main bodies (invisible on long-exposure DAFO photographs).
The fragments can be visible in even more details on the FIPS (Fireball Intelligent Positioning System) cameras installed in Ond\v{r}ejov
and Kun\v{z}ak. The cameras with a field of view of about $15\degr$ on a moving mount are tracking fireballs according 
to a navigation all-sky video. FIPS captured at least partly six fireballs and showed separated fragments in four of them
(2017\,Dec\,17, 2018\,Apr\,08, 2018\,Sep\,11, and 2018\,Oct\,08). Fragment positions could not be calibrated but  their presence qualitatively
confirmed results from other cameras and from the modeling.

Additional data outside the EN were used in some cases. An amateur
video was used for the 2018\,May\,23 fireball. One video camera of the Croatian Meteor Network was
used for both 2018\,Apr\,08 and 2018\,Nov\,29 (in both cases the video showed very end of the fireball).
Amateur still photographs were used to improve the trajectory solution for 2017\,Nov\,14, 2018\,Apr\,08, and 2018\,May\,23.

Spectra were obtained for 11 fireballs (all except the first three in Table~\ref{fireballlist}), either by a spectral DAFO or
by a supplementary video camera \citep[see][for the description of the EN spectral program]{IMC}. The spectra have 
not been yet analyzed qualitatively but are consistent with chondritic compositions of the meteoroids. The usual lines of 
Na, Mg, and Fe dominate the spectra.

Meteorite searches were performed in all 14 cases. Some of them were rather brief, some of them were
intensive -- depending on the conditions in the strewn field. No meteorites were found.

\begin{figure}
\centering
\includegraphics[width=8.5cm]{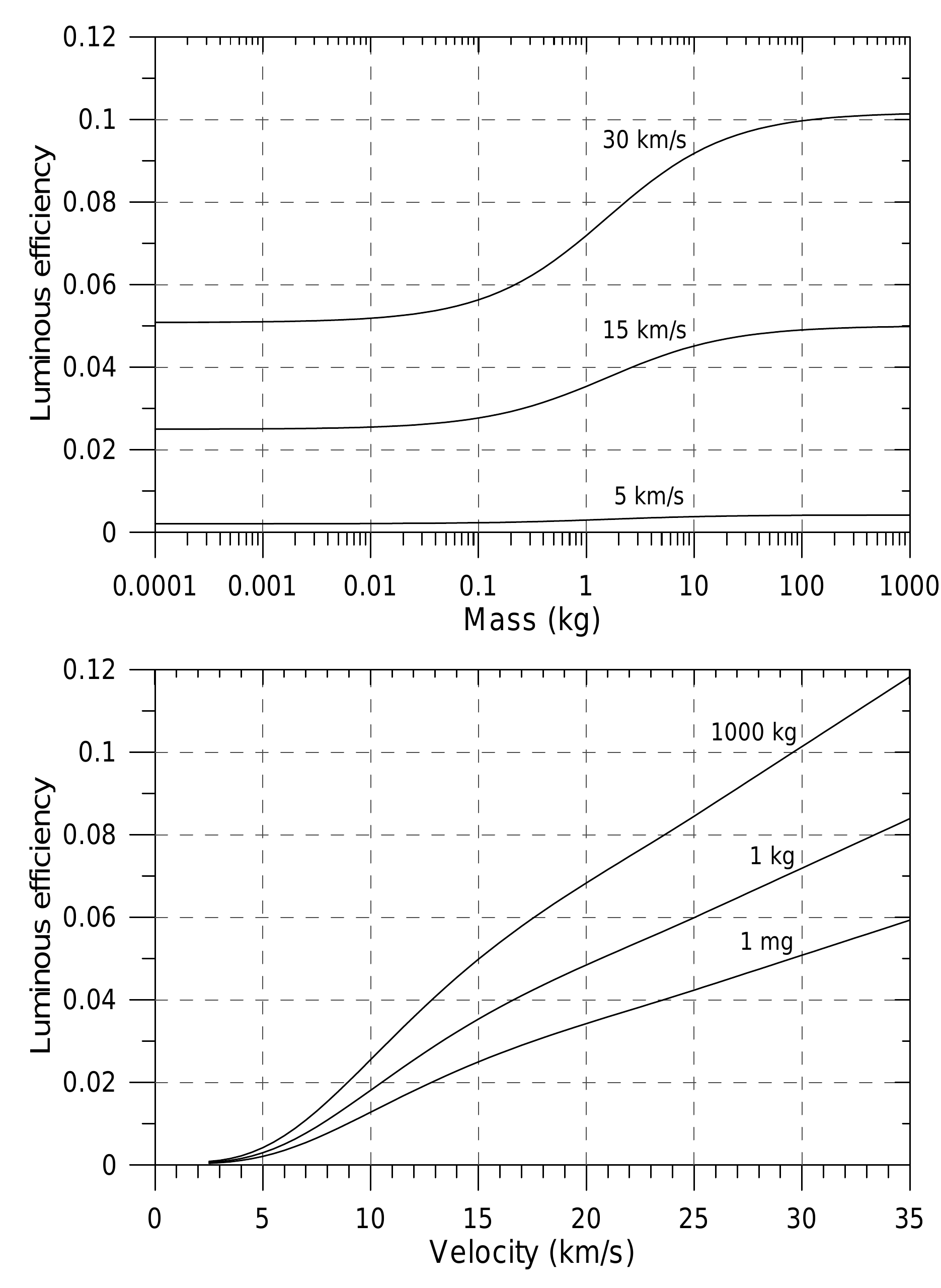}
\caption{Assumed luminous efficiency as a function of meteoroid mass and velocity.}
\label{tau}
\end{figure}

\section{The semi-empirical  fragmentation model}
\label{model}

\subsection{Description of the model}

The fragmentation model used to fit the observed data was first developed for the analysis of the
Ko\v{s}ice meteorite fall \citep{Kosice}. It has been also described in the review by \citet{chapter1}.
A more detailed description is given here. We call the model semi-empirical since the locations of 
fragmentation points must be determined from empirical data for each modeled fireball.

The model assumes finite number of fragments, which move independently.
At any time, the total luminosity of the fireball is the sum of luminosities of all individual fragments. The
basic physical theory of meteors \citep{SSR} is used to compute the motion, ablation, and radiation of
fragments. The fragment luminosity is proportional to the loss of kinetic energy:
\begin{equation}
I = - \tau(v,m) \left( \frac{v^2}{2} \frac{{\rm d}m}{{\rm d}t} + mv\frac{{\rm d}v}{{\rm d}t} \right),
\label{lumeq}
\end{equation}
where $\tau$ is the luminous efficiency (dimensionless), $v$ is the velocity and $m$ is the mass of the fragment, and
$t$ is time. 
The luminosity $I$ can be considered either in the whole spectral range or in a limited range (e.g.\ visual). The luminous efficiency
must be adjusted accordingly. In this paper we consider the whole spectral range.
The luminous efficiency is assumed to depend on velocity and mass. We used the velocity dependence
found by \citet{Pecina83} and confirmed by \citet{CepKiruna}. The mass dependency was adapted from \citet{CepKiruna}
so that the shape of the function is the same but the luminous efficiency for small meteoroids is not so low as assumed by
\citet{CepKiruna}. We believe that 
luminous efficiency may be low (of the order of 0.1\%) for small meteoroids observed by TV techniques high
in the atmosphere but for small fragments separated from bigger meteoroids lower in the atmosphere we 
found that values of about 2.5\% at 15 km s$^{-1}$ are needed. For large meteoroids ($\gg 1$ kg),  5\% at 15 km s$^{-1}$
was used. The full expression for luminous efficiency
$\tau$ (in percent) is:
\begin{eqnarray}
\ln \tau &=&  0.567-10.307\, \ln v+9.781\, (\ln v)^2-3.0414\, (\ln v)^3 + 
   \,0.3213\, (\ln v)^4+0.347\tanh(0.38 \ln m) \label{tau1} \\
&&{\rm for } \ \ v < 25.372 \nonumber  \\
\ln \tau  &=& -1.4286 +\ln v +0.347\tanh(0.38 \ln m) \label{tau2} \\
 &&{\rm for } \ \ v \ge 25.372, \nonumber 
\end{eqnarray}
where $v$ is in km s$^{-1}$ and $m$ is in kg ($\ln$ is natural logarithm and $\tanh$ is hyperbolic tangens).
The dependency is presented in graphical form in Fig.~\ref{tau}.

Since $\tau$ is dimensionless, the intensity $I$ in equation~(\ref{lumeq}) is in energetic units. 
To compare it with observations, $I$ must be converted to absolute magnitudes, $M$, by
\begin{equation}
M = -2.5 \log(I/1500),
\end{equation}
where $I$ is in watts. The equation follows from the estimate that for usual meteor plasma temperature of 4500 K, 
a zero magnitude meteor radiates 1500 W 
into the whole spectral range
\citep{SSR}.

Meteoroid deceleration and ablation are given by the drag equation and ablation equation, respectively:
\begin{eqnarray}
\frac{{\rm d}v}{{\rm d}t} &=& - \Gamma\!A \delta^{-2/3}\,\rho m^{-1/3} v^2, \\
\frac{{\rm d}m}{{\rm d}t} &=& - \Gamma\!A \delta^{-2/3}\,\sigma \rho m^{2/3} v^3,
\label{eqablation}
\end{eqnarray}
where $\Gamma$ is the drag coefficient ($0<\Gamma\leq2$), $A$ is the shape coefficient ($A=SV^{-2/3}$, where $S$ 
is head cross-section and $V$ is volume; for sphere $A=1.21$), $\delta$ is meteoroid bulk density, $\rho$ is atmospheric
density (a function of atmospheric height), and $\sigma$  is the ablation coefficient. \citet{SSR} presented integral solution
of these equations, based on the work of \citet{Pecina83}.  The solution assumes that the product $\Gamma\!A \delta^{-2/3}$
and the ablation coefficient $\sigma$ are constant. The fireball trajectory is assumed to be linear but the Earth curvature
is taken into account. The relation between the length along the trajectory and atmospheric height is therefore quadratic.

The integral solution makes it possible to compute analytically the position, velocity, mass, and luminosity of each fragment
as a function of time. The input values are the height, velocity, and mass at the initial time, the known trajectory, and the constant 
parameters $\Gamma\!A$, $\delta$, and $\sigma$. Atmospheric densities are taken from the NRLMSISE-00 model \citep{atmosphere}.
The computation proceeds until the next fragmentation point or until the velocity decreases to 2.5 km s$^{-1}$, when the ablation
and radiation are supposed to cease.

The model considers individual fragments, multiple fragments, and dust. Fragments are formed in gross-fragmentation events.
Dust can be released either suddenly or by erosion. Multiple fragments are simply identical fragments (of the same mass and
all other parameters), formed at the same time. The computation is then performed only once and the resulting
luminosity is multiplied by the number of fragments. Of course, this is an idealization intended to save the computation
time. Similarly, dust is a large number of usually small fragments (dust particles) in a given mass range. 
The masses of dust particles are sorted into mass bins. The number of mass bins per order of magnitude of mass,
$b$, can be chosen. For example, if $b=4$, the logarithms of masses are separated by 0.25.
The parameters of the dust are the total mass, $D$, the upper and lower mass limits
of dust particles, $m_0$ and $m_k$, respectively, and the mass distribution index, $s$. A power law mass distribution
is assumed.  The number of particles in the $i$-th bin with masses $m_i$ is
\begin{equation}
n_i=n_0 \left(\frac{m_0}{m_i} \right)^{s-1}, \ \  (i=0,1,\dots,k), 
\label{eqni}
\end{equation}
where $k$ is number of additional mass bins (i.e.\ other than the bin containing the particles of the largest mass),
\begin{equation}
k=b \log(m_0/m_k),
\label{eqk}
\end{equation}
and
\begin{equation}
m_i = m_0\,10^{-i/b},  \ \ (i=0,1,\dots,k).
\end{equation}
The number of particles of the largest mass is 
\begin{equation}
n_0 = \frac{D}{(k+1)m_0}, \ \ {\rm for } \ \ s = 2, 
\label{eqn0-1}
\end{equation}
or
\begin{equation}
n_0 = \frac{D}{m_0}\cdot\frac{1-p^{(2-\!s)}}{1-p^{(k+1)(2-\!s)}}, \ \  {\rm for } \ \ s \neq 2, 
\label{eqn0-2}
\end{equation}
where $p = 10^{-1/b}$. Since $n_i$ and $k$ are integer numbers, integer parts are taken in equations
(\ref{eqni}), (\ref{eqk}), (\ref{eqn0-1}), and (\ref{eqn0-2}).

\begin{figure}
\centering
\includegraphics[width=8.5cm]{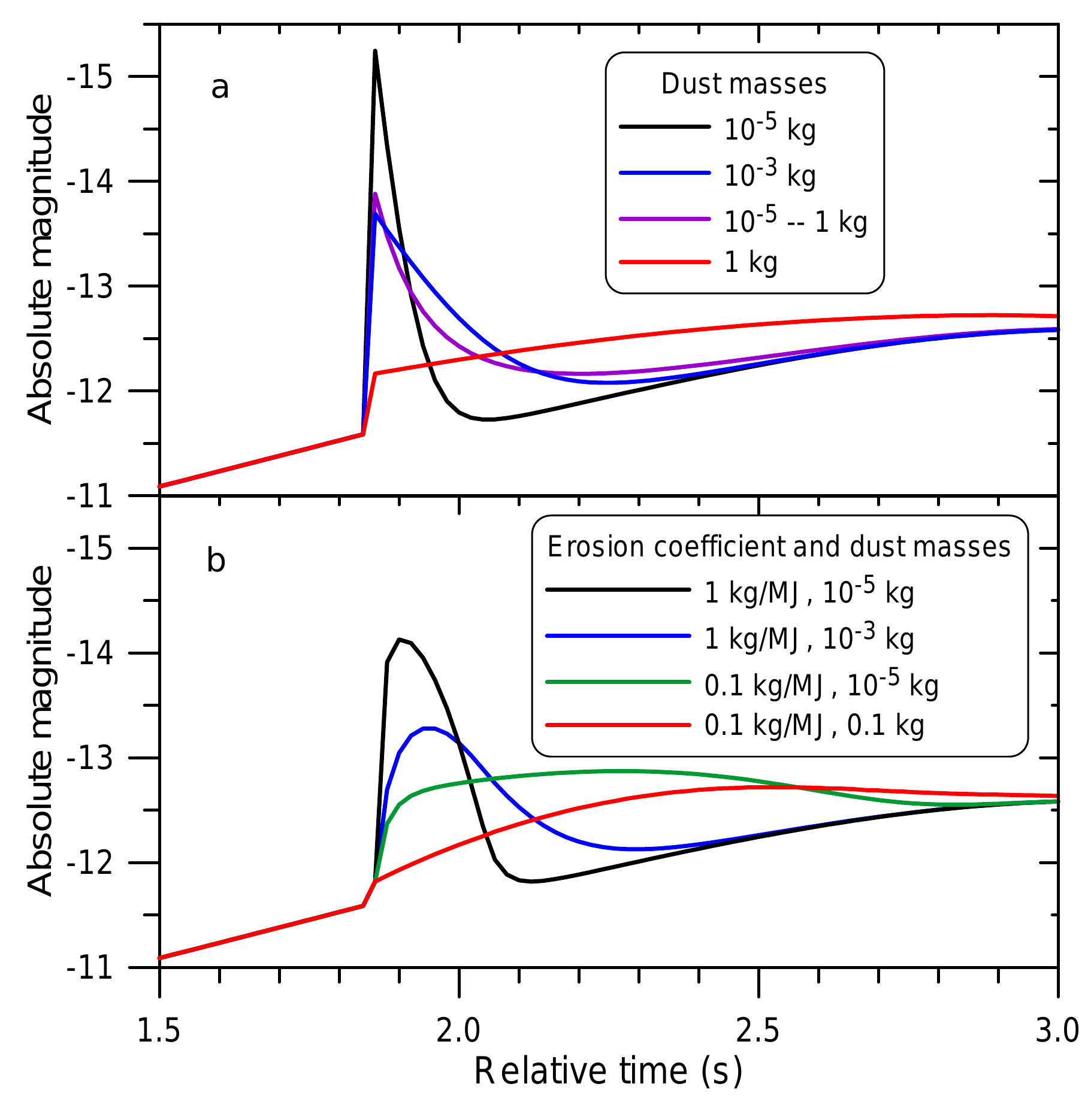}
\caption{Effects of single fragmentation events with various parameters on shape of the light curve. Sudden releases
of dust or multiple fragments (a) and formations of an erosion fragment (b) are shown. In all cases, a 100 kg meteoroid moving
at 15 km s$^{-1}$ at the height of 40 km looses 25 kg of mass. The other parameters are $\Gamma\!A=0.8$, 
$\delta=3400$ kg m$^{-3}$, $\sigma=0.005$ kg MJ$^{-1}$ for all fragments and dust particles, $s=2$, and trajectory slope
45\degr. Computations are done with the time resolution 0.02 s.}
\label{lceffects}
\end{figure}

The effects of a sudden dust release on the light curve is shown in Fig.~\ref{lceffects}a. The release of small dust particles
produces a short bright spike. Larger dust particles produce longer and asymmetric peaks with quick rise and slower decay.
The formation of multiple large fragments produce a step on the light curve.

To be able to reproduce more symmetric peaks and long humps often observed on real light curves, the concept of 
eroding fragments was introduced. The erosion formalism is the same as formulated for faint Draconid meteors by
\citet{Draco}. The concept of quasi-continuous detachment of small particles was, nevertheless, formulated much
earlier \citep[e.g.][]{Simonenko}. In our approach, 
the eroding fragment is loosing mass by ablation described by equation~(\ref{eqablation})
and by erosion described by an analogical equation where the ablation coefficient $\sigma$ is replaced by the
erosion coefficient $\eta$. Both coefficients have the same unit, kg MJ$^{-1}$ or, equivalently, s$^2$km$^{-2}$.
The ablated mass is in the vapor form and immediately contributes to fireball radiation. The eroded mass is released
in form of dust particles, which only subsequently ablate and radiate. The dust parameters are $m_0$, $m_k$, and $s$
as in the case of immediately released dust. The dust mass distribution is computed from equations (\ref{eqni}) -- (\ref{eqn0-2}),
where the total dust mass $D$ is replaced by the mass eroded within the time step of computation.

The light curve effects produced by eroding fragments are illustrated in Fig.~\ref{lceffects}b for four combinations of the
erosion coefficient $\eta$ and the mass of dust particles. In these examples, all dust particles were supposed to have the same mass.
The erosion continued until the eroding fragment was completely exhausted. The model formalism allows to stop the erosion
(and continue only ablation) after a prescribed part of the fragment was eroded out, but this feature was not used in modeling
the fireballs presented here. Also, the eroding fragments were not subject to further gross-fragmentation events. Individual
fragments and multiple fragments, on the other hand, could fragment repeatedly. Dust particles, released either suddenly or by erosion, 
cannot be subject to fragmentation in this model.

\subsection{Modeling procedure}

The model was used to fit the observed fireball light curves and dynamics, i.e.\ length along the trajectory as a function of time.
The trajectory was always known from multi-station linear least squares solution \citep{BAC1990}. 
A point at the trajectory near the observed fireball beginning
was chosen as the starting point. Here the meteoroid was assumed to be a single body described by its mass, density,
velocity, drag, shape, and ablation coefficients. The velocity was tuned to fit the observed dynamics at the beginning of the fireball. 
The first guess of the mass was obtained from the total radiated energy, taking into account the velocity and the approximate
luminous efficiency for that velocity. The initial mass had to be, nevertheless, adjusted when the modeling proceeded.
The density of corresponding meteorites was used for meteoroid density. If no meteorite was recovered, a value of 3400 kg m$^{-3}$ was
assumed. For other parameters, the canonical values were $\Gamma\!A=0.8$ and $\sigma=0.005$ kg MJ$^{-1}$.
In order to keep the number of free parameters of the model at minimum, these parameters and the density were used, if possible, also
for all subsequent fragments and dust particles. Such a uniform approach to all fireballs was preferred not to
introduce biases when studying meteoroid fragmentation behavior, which was the main goal of this work.

\begin{figure}
\centering
\includegraphics[width=14cm]{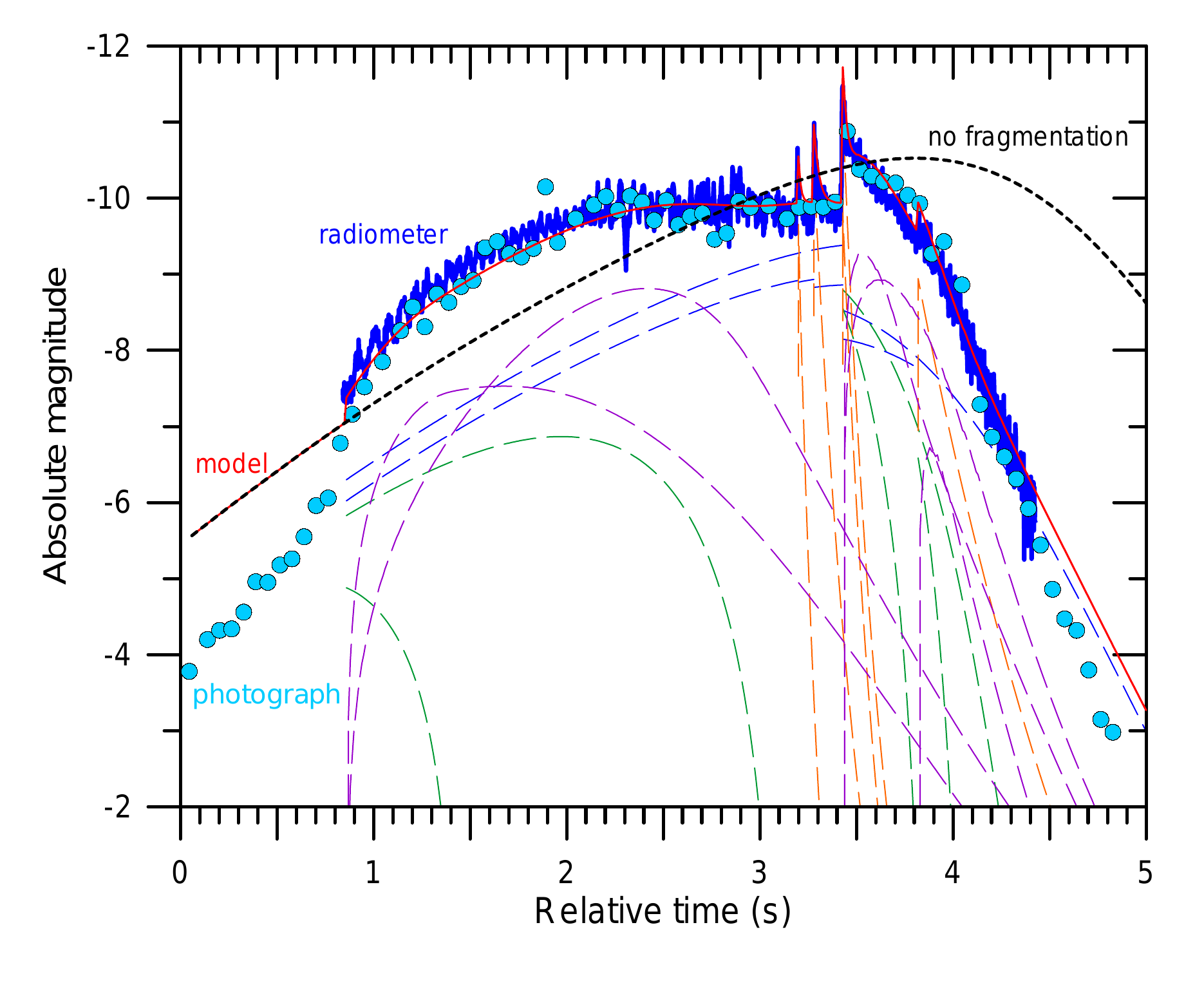}
\caption{Observed and modeled light curve of fireball 2015\,Aug\,26. The radiometric light curve (blue line) and photographic data 
(light blue dots) are shown from only one station for clarity. The modeled light curve is shown as solid red line. The
contributors to the light curve are shown as dashed lines: blue for regular fragments, green for eroding fragments,
orange for immediately released dust and violet for dust from eroding fragments. The light curve modeled without any fragmentation
is shown as dotted black line for comparison.}
\label{lcexample}
\end{figure}

The modeling was done manually by a trial-and-error method. Fragmentation points were identified according to features
on the light curve (flares, sudden changes in slope). Another sign of fragmentation can be a sudden increase of deceleration. 
In few cases, fragmentation point was identified according to a change of trajectory, i.e.\ change of direction of flight (by 1--2\degr),
before the end of the fireball. The change was supposed to be produced by a lateral impulse acquired during break-up. 
Since there was no flare and no big change of deceleration in the observed cases, 
the fragmentation was modelled by a loss of relatively small
amount of mass in the form of few fragments. Such cases demonstrate that there may be some fragmentation events
which are not apparent in the light curve or dynamic data and can remain unnoticed if there is no sufficiently large lateral impulse
or geometric data are not precise enough.
On the other hand, these were only minor events with minor mass loss.

As an example, light curve modeling is shown in Fig.~\ref{lcexample} for fireball 2015\,Aug\,26. The initial meteoroid mass
was found to be 6.5 kg. At the beginning, the modeled
brightness was higher than observed, since the model assumes a steady-state ablation,  while in reality, the ablation started gradually. 
At time $\approx 0.8$~s, the brightness started to increase rapidly.
A hump was formed on the light curve lasting for two seconds (time 1 -- 3~s). The brightness was elevated above the level expected for a single
non-fragmenting body. This feature could be modeled by a disruption of the 6.5 kg meteoroid into two regular (2.6 kg and 1.5 kg)
and two eroding (1.9 kg and 0.5 kg) fragments at the time 0.9 s (height 69 km, dynamic pressure 0.04 MPa). 
The ablating dust released gradually from the eroding fragments formed the hump. In fact, each ``dust'' particle was assumed
to have mass of one gram here. The erosion coefficients of the two eroding fragments were 0.1 and 
1~kg MJ$^{-1}$, respectively. Of course, these parameters are only schematic. The substantial fact is the identification
of the fragmentation at the height of 69 km and the fact that the following two seconds of the flight could be fitted without further
disruption (erosion is considered here as thermal process analogous to ablation, not as a ``real'' fragmentation).
Note that there are semi-regular oscillations visible on the radiometric light curve (confirmed on several independent radiometers),
which are not attributed to fragmentation. It might be a demonstration of an instability process in ablation.

\begin{figure}
\centering
\includegraphics[width=8.5cm]{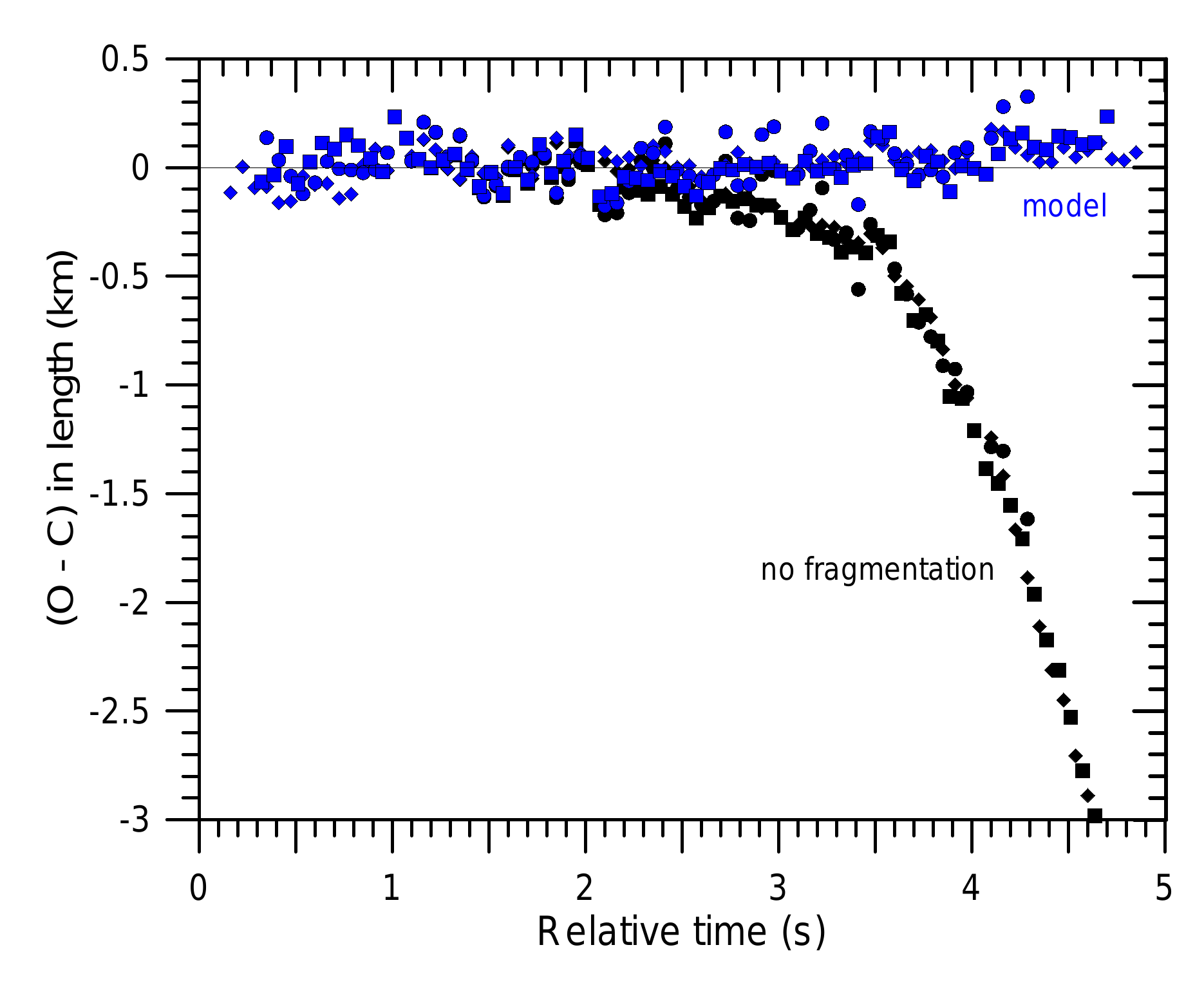}
\caption{Dynamics of fireball 2015\,Aug\,26. The difference between the observed and modelled length along the trajectory is
shown as a function of time. Black symbols are for model without fragmentation; blue symbols for the fragmentation model.
The small positive trend of the blue symbols at the end was allowed to account for gravity acceleration, which was not modeled directly.
Different symbols correspond to observations from different stations.}
\label{dynexample}
\end{figure}

The initial fragmentation at 0.9 s can be confirmed also from fireball dynamics. Dynamics also
enabled us to determine the mass of the largest fragment after the fragmentation. The observed length along the trajectory
deviates from that expected for non-fragmenting body already at time 2.5 s (Fig.~\ref{dynexample}). The dynamics between
0.9 s and 3.4 s corresponds to a fragment of initial mass 2.6 kg, decreasing during that interval to 2.1 kg due to ablation. The
masses of other modelled fragments were estimated from the light curve fit. Unless there is a video record showing more
fragments, dynamic data are available only for the leading (foremost) body. Note that the leading bodies can exchange, as it was
directly observed in videos of Mor\'avka \citep{Moravka} and Chelyabinsk \citep{Celjab} superbolides.

After more than two seconds of quiescence, the 2015\,Aug\,26 fireball showed three short flares at 3.2, 3.3, and 3.4 s
(heights 43 -- 40.5 km, dynamic pressures 0.9 -- 1.15 MPa). After the third flare (the largest one), the brightness remained
elevated for about 0.3 s and then decreased rapidly, especially after another, smaller, flare at 3.8 s (37 km, 1.4 MPa). 
The flares at 3.2 and 3.3 s were modelled as immediate dust releases of particles of $10^{-5}$ -- $10^{-6}$ kg (mm-sized).
For the flare at 3.4 s, two eroding fragments were added. The eroded dust formed the bulk of radiation at 3.5 -- 3.8 s. 
The particle masses were $10^{-4}$  -- 0.05 kg (cm-sized); the erosion coefficients were 0.05 -- 0.1 kg MJ$^{-1}$. 
Similar erosion parameters were used for the last flare at 3.8 s, where no mm-sized dust was released. 

The fragmentation
sequence, i.e.\ which fragment fragmented at which time, cannot be revealed unambiguously. But the quick increase
of deceleration in comparison with the no-fragmentation case at time 3.5 s (Fig.~\ref{dynexample}) shows that the
mass of the leading fragment decreased significantly. To model the dynamics and light curve, it was assumed that
two 0.5 kg fragments emerged after the main fragmentation at 3.4 s. One of them was destroyed at 3.8 s.
For the second fragment, no further fragmentation was needed to fit the dynamics, but the ablation coefficient had to be
enhanced from the nominal value of 0.005 kg MJ$^{-1}$ to 0.01 kg MJ$^{-1}$. The mass of this fragment (0.5 kg when
formed and 0.12 kg at the end) was determined from the dynamics. 
In addition to this single meteorite, some meteorites smaller than 0.03 kg could result from the largest bodies modeled as dust.
The modeled brightness toward the end of the fireball was somewhat higher
than observed (Fig.~\ref{lcexample}). This discrepancy could be removed either by decreasing the luminous efficiency
at speeds below 10 km s$^{-1}$ or by lowering simultaneously both the mass and $\Gamma\!A$ of the fragment.

The example of the 2015\,Aug\,26 fireball demonstrates the interplay between light curve and dynamics. The positions
of the important fragmentation points can be determined robustly from the light curve. The initial mass of the meteoroid
and the amount of mass released as dust is also determined from the light curve (and known velocity). The masses
of leading fragments can be determined from fireball dynamics, i.e.\ observed deceleration. Other fragments
can be studied directly only if they were imaged on a video. Otherwise, the evidence of their existence is obtained only
indirectly from their contributions to the light curve 
and the adopted solution (number and masses of fragments) is usually not unique.
Unless some fragments are well observed from two (or more) 
well separated stations, all fragments are assumed to follow the same trajectory as the main body. Of course, all
derived masses depend on the assumed values of luminous efficiency, fragment density, shape etc. Masses are therefore less
certain than the dynamic pressures at fragmentations.

\begin{figure}
\centering
\includegraphics[width=18cm]{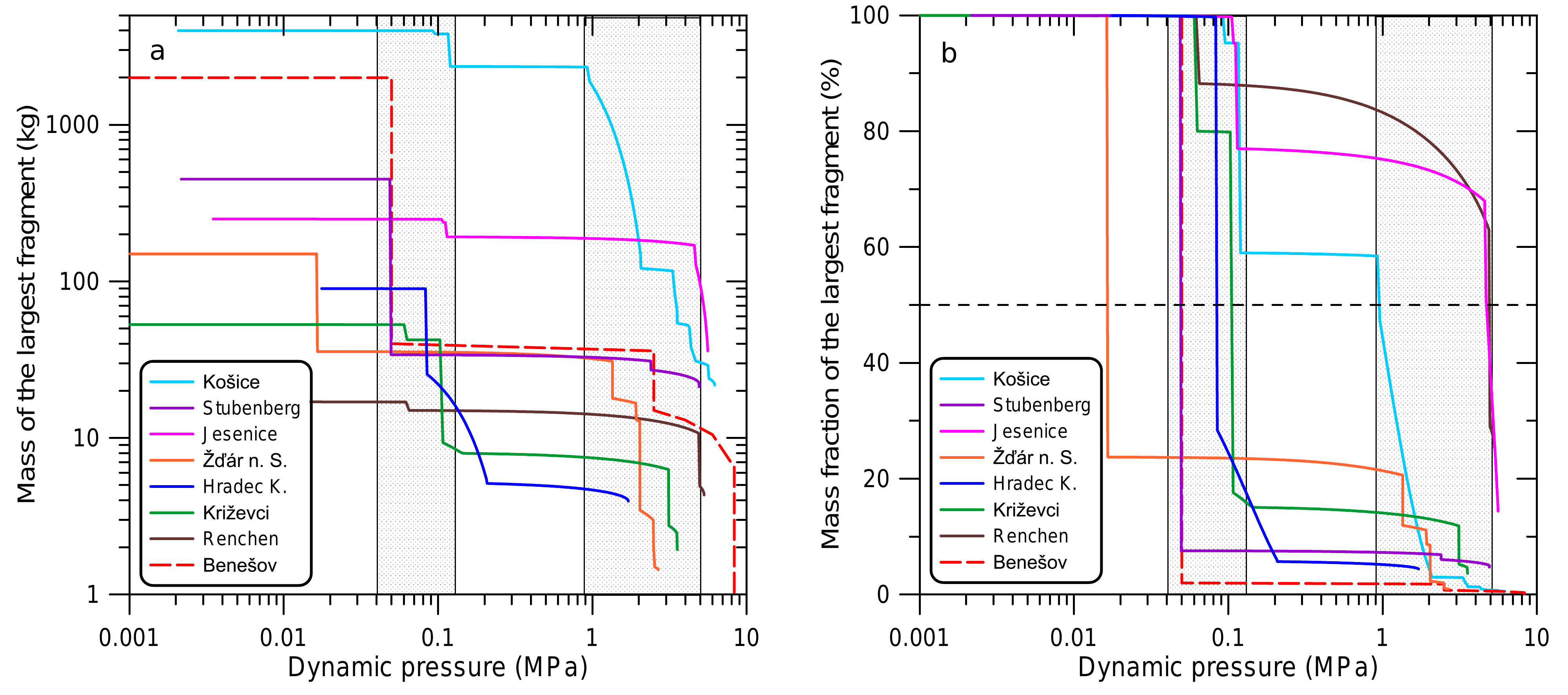}
\caption{Modeled mass of the largest surviving fragment as a function of increasing dynamic pressure for seven
meteorite falls. Absolute mass values are plotted in panel {\textbf a}; percentages of the initial mass are plotted in panel {\textbf b}.
The result of an approximate analysis of the Bene{\v s}ov meteorite fall of 7 May 1991 \citep{IAU318} is included
for comparison. The fragmentations preferentially occurred in two dotted intervals of dynamic pressures.
One exception was the early fragmentation of \v{Z}\v{d}\'ar nad S\'azavou.}
\label{p-m1}
\end{figure}

\section{Results}
\label{results}

The procedures explained in Section~\ref{model} were applied to the observations of verified and suspected meteorite falls
described in Section~\ref{instrum}. The results are presented here separately for both groups.

\subsection{Meteorite falls}
\label{results_meteorites}

In Fig.~\ref{p-m1}a, the results of fragmentation modeling of seven meteorite falls from Table~\ref{meteoritelist}
are presented. The approximate analysis of the Ben\v{s}ov meteorite fall from \citet{IAU318} is added for comparison.
Note that the models for Jesenice and
Kri\v{z}evci were slightly revised in comparison with original publications. For Jesenice, the fragmentation 
at the height of 46 km was omitted since it was based on seismic data, which are not very reliable for these heights,
and there is no sign of fragmentation at this height in the light curve.
The light curve suggests an earlier fragmentation at about 55 km.

The mass of the largest surviving fragment is plotted as a function of dynamic pressure. Dynamic
pressure always increases along the trajectory as the meteoroid penetrates into denser atmospheric layers. Only when the
meteoroid is decelerated significantly toward the end of the luminous trajectory, the dynamic pressure
starts to decrease. The part of the trajectory with decreasing pressure is not plotted. Although the fragmentation
can continue here and also during the dark flight, 
as it is evidenced by incomplete fusion crust of some meteorites \citep[e.g.][]{Moravka, Bischoff_Renchen},  the
reason is not increasing dynamic pressure. Probably it is an aftermath of previous fragmentations.

Figure~\ref{p-m1}a shows step-wise decreases of masses due to gross-fragmentations, gradual but steep decreases
due to erosion (e.g.\ for Hradec Kr\'alov\'e) and only slight decreases due to ablation. If the erosion decrease
was followed by a much slower decrease, it does not mean that the erosion stopped at that point. It means that the
mass of the eroding fragment decreased below the mass of a regular fragment, which then became the largest 
surviving fragment.

It is obvious from Fig.~\ref{p-m1}a that the gross-fragmentations did not occur randomly. 
We can clearly see two phases of fragmentations.
The first phase occurred typically between 0.04 -- 0.12 MPa.  Only \v{Z}\v{d}\'ar nad S\'azavou fragmented
already at 0.017 MPa. The second phase of fragmentation occurred at 1 -- 5 MPa (for Ko\v{s}ice at 0.96 - 5.7 MPa).
No gross-fragmentation events occurred in these seven fireballs at intermediate dynamic pressures of
0.12 -- 0.96 MPa.

Figure Fig.~\ref{p-m1}b shows the same data but the mass of the largest fragment is normalized to the
initial meteoroid mass. In four of the seven cases, the first phase of fragmentation can be considered as catastrophic.
Disruption was defined as catastrophic (in other context -- tidal disruption of asteroids) when the largest surviving 
fragment contains less than 50\% of the original mass \citep{Richardson}. For Ko{\v s}ice, the first phase was also almost
catastrophic, with surviving fragment mass of about 60\% of the original mass. In fact, the mass of the surviving fragment 
is not well restricted by the
data in this case. Replacing the $\approx 2400$ kg fragment in the model by two pieces of half mass 
would be also consistent with observations. So, only Renchen and Jesenice were surely not disrupted catastrophically in the first
phase. They lost less than 25\% of mass there. Both these meteoroids were also most resistant in the second stage
of fragmentation. They were disrupted only at about 5 MPa.

The second stage of fragmentations was severe in most cases. Only Hradec Kr\'alov\'e, which was disrupted 
into relatively large fragments during the first phase and during the subsequent erosion process, showed no significant
fragmentation above 1 MPa -- at least judging from the light curve. There are no dynamic data at the end of the
trajectory. The recovered meteorite showing only thin fusion crust on part of the surface, nevertheless, demonstrates that there was
a late stage fragmentation \citep{Spurny_MB106}.

\begin{figure}
\centering
\includegraphics[width=16cm]{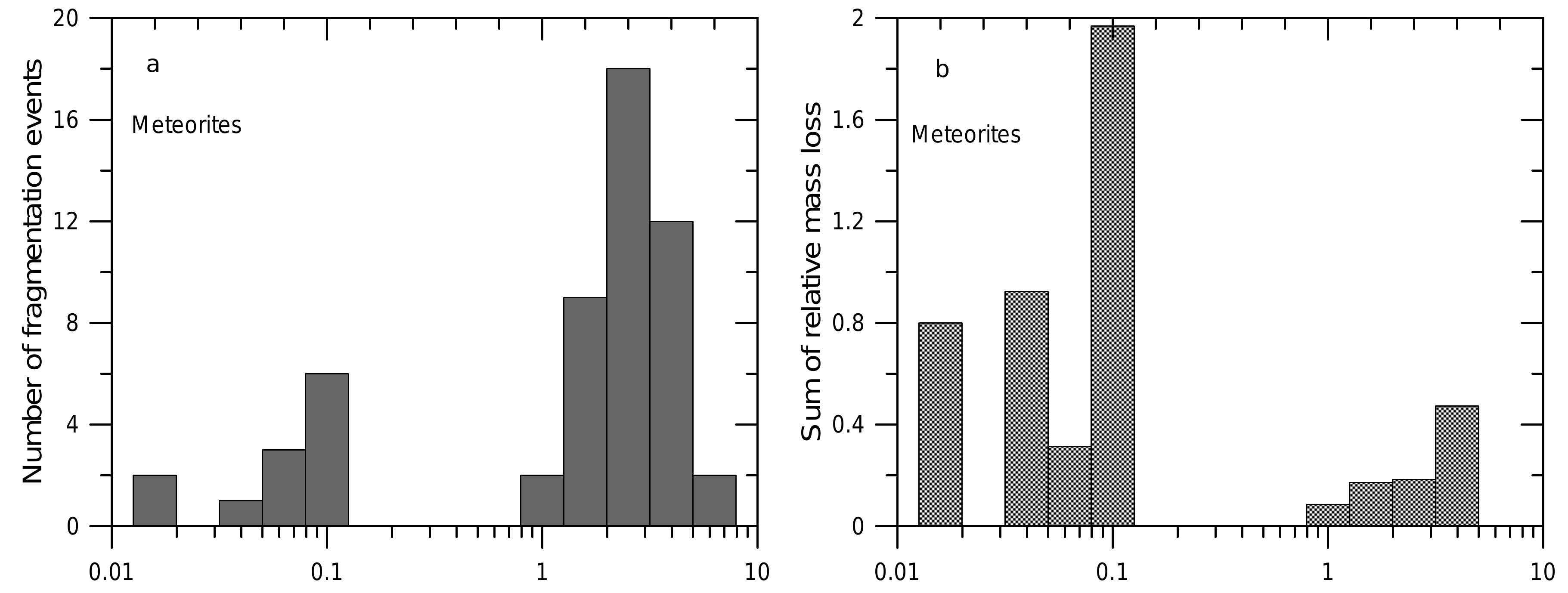}
\includegraphics[width=16cm]{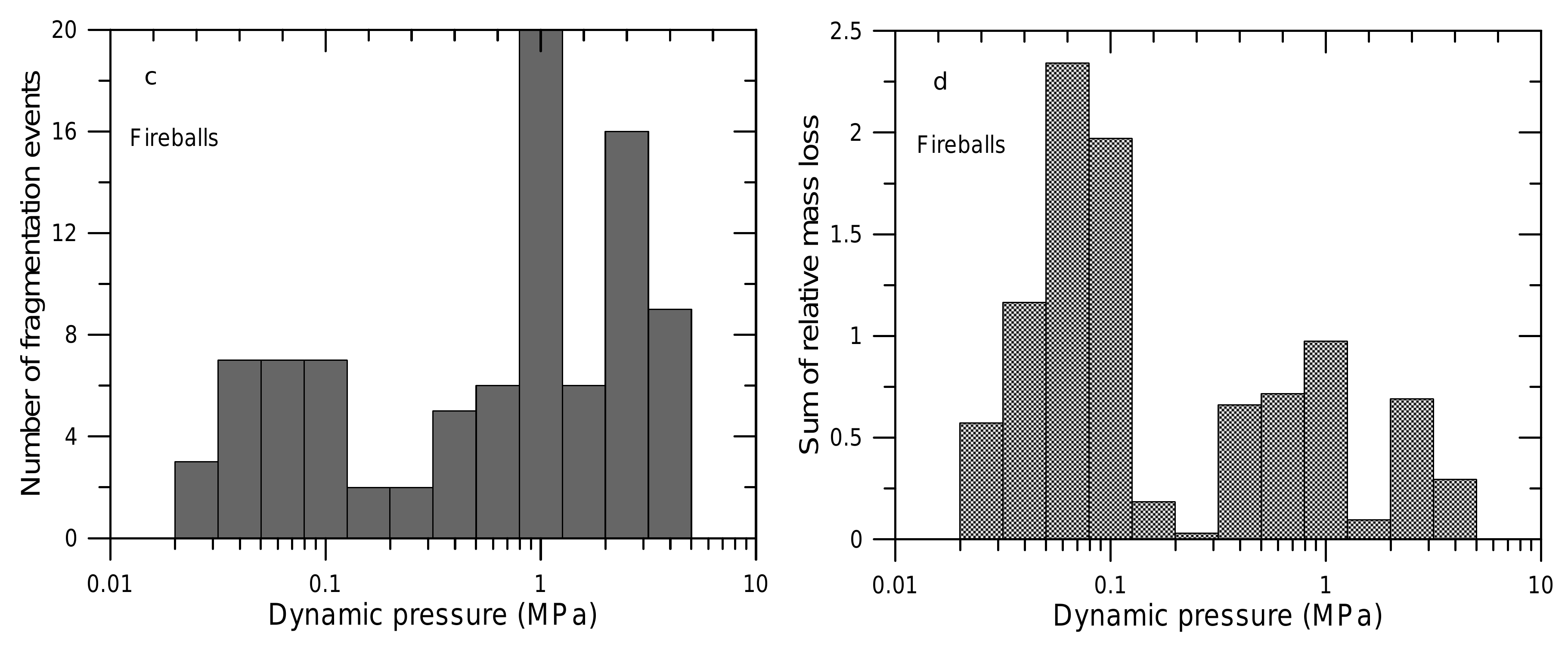}
\caption{Histograms showing the distribution of dynamic pressures at gross-fragmentation points. Panels {\textbf a} and
{\textbf b} are for recovered meteorite falls, panels {\textbf c} and {\textbf d} for fireballs with unrecovered meteorites.
Panels {\textbf a} and {\textbf c} show the number of fragmentation events. In panels {\textbf b} and
{\textbf d}, the relative mass losses are summed (see the text for explanation). }
\label{histog1}
\end{figure}

Figure~\ref{p-m1} shows only the fragmentations that occurred to the largest fragment at the given time. 
We therefore present in 
histograms in Fig.~\ref{histog1}a-b all gross-fragmentation events in all seven meteorite
falls. Panel {\textbf a} shows simply the number of events in each interval of dynamic pressures. In panel
{\textbf b}, each fragmentation event was weighted by the relative mass loss. Mass loss is defined here as the difference between
the mass before fragmentation and the mass of the largest regular (i.e.\ non-eroding) fragment after the fragmentation.
Relative mass loss is defined as the mass loss divided by the initial mass of the whole meteoroid at the atmospheric entry.

The dichotomy of fragmentation strengths is clearly visible also in Fig.~\ref{histog1}a-b.  Most of the mass was lost
in the first phase of fragmentation at pressures up to 0.1 MPa. The fragmentations in the second phase were more 
numerous but involved less mass.  The reason was that the first phase usually produced a number of small
fragments, which then disrupted in the second phase but the mass loss was small in each case. 
Some fragments lost mass repeatedly in small amounts during the second phase. Only Jesenice
and Renchen lost most mass during the second phase. There was no case with negligible fragmentation so that
most mass would be lost just by ablation.  In that case, the meteorite could contain significant part 
of the initial mass. Such cases exists but are rare. One example was the Carancas crater forming event
\citep{Carancas, Brown_Carancas}.

\begin{figure}
\centering
\includegraphics[width=18cm]{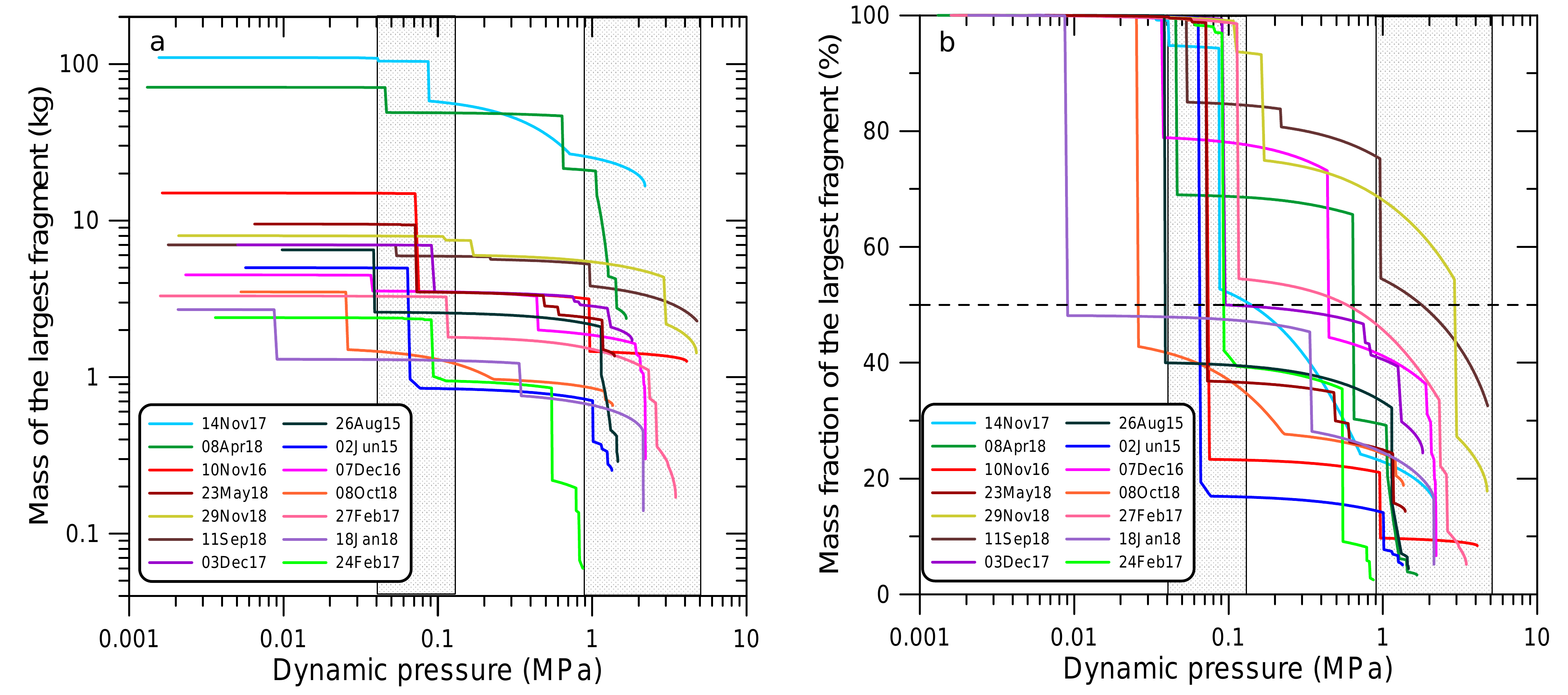}
\caption{The same as in Fig.~\ref{p-m1} for 14 fireballs classified as ordinary chondrite meteorite falls  
but where meteorites remained unrecovered.}
\label{p-m2}
\end{figure}

\subsection{Fireballs}

The results of analysis of meteorite-dropping fireballs with unrecovered meteorites are presented 
in Fig.~\ref{histog1}c-d and Fig.~\ref{p-m2}. Except for two cases, these fireballs were produced
by small meteoroids of masses not exceeding 20 kg. Smaller mass is, in fact, an advantage
for studying the outcome of the first phase of fragmentation. Since smaller meteoroids show deceleration
already at middle heights corresponding to dynamic pressures of tenths of MPa, it is 
possible to compute the mass of the largest fragment at these heights from dynamics.

Fig.~\ref{histog1}c-d shows that the bimodality of strengths is present in these fireballs as well, though
in contrast to cases with recovered meteorites, fragmentation events were observed also at
pressures 0.1 -- 1 MPa. There were only few minor fragmentations at 0.12 -- 0.3 MPa. 
More significant events occurred at 0.3 -- 1 MPa. Figure~\ref{p-m2} shows that major fragmentations
of the leading fragments occurred at these pressures in four fireballs. Two of them are the two
smallest meteoroids in our sample, 2018\,Jan\,18 and 2017\,Feb\,24, both with mass less than 3 kg at the entry,
i.e. of a diameter of about 11 cm. 

The third fireball that fragmented under a medium pressure was 2016\,Dec\,07. This fireball was in fact the most difficult to model. 
It was not possible to fit simultaneously the light curve and deceleration with the nominal set of parameters.
There is a clear indication of fragmentation on the light curve at the height of 48 km, under the pressure of 
0.45 MPa. The deceleration after fragmentation corresponds to a fragment mass of $\sim 8$ kg for
the nominal values of $\Gamma\!A=0.8$ and $\delta=3400$ kg m$^{-3}$. However, the fireball
brightness corresponds to a mass of less than 2 kg. The discrepancy was formally solved by assuming
very low $\Gamma\!A$ for some fragments, down to 0.45. It would physically mean elongated shape
and orientation with the smallest cross-section in the direction of flight. Alternatively, a higher than nominal density
(combined with less elongated shape) or a lower luminous efficiency could be assumed. Such dramatic
changes of nominal values were not needed for any other fireball. We have a suspicion that this meteoroid
was not an ordinary chondrite. The spectrum of this fireball does not look exceptionally and clearly excludes 
the meteoroid being an iron-nickel.
The brightest lines belong to sodium and magnesium.

The fourth fireball with medium-pressure fragmentation was 2018\,Apr\,08. It was caused by a relatively
large meteoroid with initial mass  $\sim 70$ kg (diameter $\sim 34$ cm). A fragmentation at 0.65 MPa
is well documented on the light curve and by one fragment seen on video. Other fragmentations 
then occurred at 1.1 and 1.4 MPa.

Generally, the analyzed fireballs document that the second stage fragmentation can start already at 0.5 MPa
or even earlier, especially for smaller meteoroids. The fact that the
second stage occurred above 0.9 MPa for events with recovered meteorites can be partly a selection effect. 
More resistant meteoroids, which
fragment later, can produce more meteorites or larger meteorites with higher chances for recovery. 
Nevertheless, even in the sample with no meteorites, the majority of severe second phase fragmentations occurred
above 0.9 MPa (Fig.~\ref{p-m2}). Note that the largest meteoroid in this sample, 2017\,Nov\,14, showed no
second stage fragmentation. The last gross-fragmentation occurred during the first phase at 0.09 MPa,
where one slowly eroding and one regular fragment were produced. The regular fragment seems to survive
pressures up to 2.2 MPa without further fragmentation and probably produced one large meteorite with a mass of about
10 kg. The shallow trajectory with the slope of only 15\degr\ to the horizontal was the reason why higher dynamic pressures
were not reached and the fragment was gently decelerated. This fireball lasted for 15 seconds.

The first phase fragmentation occurred in all 14 fireballs. It was catastrophic with $\gtrsim60$\% mass loss 
for six of them and nearly catastrophic with $\sim50$\% mass loss for other four. Only four fireballs showed $\lesssim30$\%
mass loss during the first phase fragmentation (see Fig.~\ref{p-m2}b). The first phase occurred, with
three exceptions, during the same range of dynamic pressures as for most recovered meteorite falls, 
i.e.\ 0.04 -- 0.12 MPa. The small meteoroid 2018\,Jan\,18 fragmented already at 9 kPa and the rather small meteoroid 2018\,Oct\,08 fragmented at 0.026 MPa.
On the contrary, 2018\,Nov\,29 finished the first phase of fragmentation at 0.17 MPa. What is, however important,
is that the two phases were well separated in almost all cases. After the first phase, there was a
quiet period when the dynamic pressure increased more than 10 times (in 10 fireballs) or at least 5 times
(in 3 fireballs) without further gross-fragmentations. Only 2018\,Sep\,11 showed a very minor mass loss in between.

The light curve of fireball 2015\,Aug\,26 which was shown in Fig.~\ref{lcexample} is rather typical. The fireballs typically exhibit
a sudden increase of brightness at the beginning followed by a hump not expected for single body but without further
strong irregularities. Only much later, flares accompanying the second phase of fragmentation occur. When
imaged on video, the fireballs also show typical changes of morphology. Long wake develops in the first half of the
trajectory. The wake then disappears and the bolide again becomes point-like. Toward
the end, a short wake is formed, which then separates into individually moving fragments. These changes are consistent
with two phases of fragmentation. The initial wake must be formed by small dust particles. Large fragments 
move together for some time since their deceleration is negligible
at high altitudes. They can be separated at lower heights, where also disruptions of the second phase occur.
A more detailed investigation of the initial wakes revealed that they indeed form at the time when fragmentation
is indicated by the light curve \citep{Shrbeny}.

\begin{table*}
\caption{Geocentric radiants and heliocentric orbits of fireballs analyzed in this study (J2000.0). 
Tisserand parameter relatively to Jupiter and initial mass from the model are also given.}
\label{orbitlist}
\begin{flushleft}
\vspace{-2em}
\begin{tabular}{rrrrllllrrrrr}
\hline \hline
Name      & \multicolumn{1}{c}{$\alpha_{\rm G}$}   & \multicolumn{1}{c}{$\delta_{\rm G}$}    &\multicolumn{1}{c}{$v_{\rm G}$} 
  & \multicolumn{1}{c}{$a$}   & \multicolumn{1}{c}{$e$}     & \multicolumn{1}{c}{$q$}     & \multicolumn{1}{c}{$Q$}     & 
\multicolumn{1}{c}{$i$}      & \multicolumn{1}{c}{$\omega$}   & \multicolumn{1}{c}{$\Omega$}      & $T_{\rm Jup}$      & Mass    \\
 & \multicolumn{1}{c}{\degr} & \multicolumn{1}{c}{\degr} & km/s & \multicolumn{1}{c}{au}& & \multicolumn{1}{c}{au} &\multicolumn{1}{c}{au} &
\multicolumn{1}{c}{\degr} & \multicolumn{1}{c}{\degr} & \multicolumn{1}{c}{\degr} &&kg\\
\hline
2015\,Jun\,02 &232.70 &26.37 &11.88 &1.794 &0.471 &0.9489 &2.64 &13.69 &216.95 &71.842 &3.91 &5\\[-0.5ex]
        &  $\pm0.02$ &0.03     &0.07 &0.012 &0.004 &0.0002 &0.02 &0.08 &0.04 & &0.02 \\

2015\,Aug\,26 &314.77 &$-$6.99 &16.00 &2.668 &0.6986 &0.8041 &4.53 &4.496 &239.77 &153.207 &2.97 &6.5\\[-0.5ex]
               &0.02   &0.02     &0.03 &0.012 &0.0015 &0.0004 &0.03 &0.016 &0.03 & &0.01\\

2016\,Nov\,10 &43.53 &66.513 &20.82 &1.282 &0.4691 &0.6804 &1.883 &30.46 &268.84 &227.891 &4.82 &15\\[-0.5ex]
               &0.06 &0.006  &0.06   &0.003 &0.0015 &0.0004 &0.007 &0.09 &0.09 & &0.01\\

2016\,Dec\,07 &78.56 &16.82 &18.17 &1.432 &0.594 &0.5821 &2.283 &3.855 &95.61 &75.289 &4.47 &4.5\\[-0.5ex]
               &0.03 &0.03  &0.07 &0.004 &0.002 &0.0013 &0.010 &0.003 &0.06 & &0.01\\

2017\,Feb\,24 &145.75 &17.93 &13.65 &1.736 &0.538 &0.8027 &2.67 &1.622 &242.70 &336.106 &3.97 &2.4\\[-0.5ex]
               &0.03 &0.03    &0.07     &0.011 &0.003 &0.0005 &0.02 &0.016 &0.05 & &0.02 \\

2017\,Feb\,27 &171.656 &6.362 &29.39  &2.249 &0.8421 &0.3552 &4.143 &3.067 &293.78 &338.442 &3.02 &3.3\\[-0.5ex]
               &0.006 &0.008 &0.03    &0.007 &0.0006 &0.0003 &0.014 &0.012 &0.02 & &0.01\\

2017\,Nov\,14 &36.86 &2.04 &16.00 &2.59 &0.694 &0.7913 &4.39 &5.419 &59.21 &52.284 &3.02 &110\\[-0.5ex]
               &0.04 &0.05 &0.07 &0.03 &0.004 &0.0005 &0.06 &0.007 &0.02 & &0.02\\

2017\,Dec\,17 &40.34 &66.802 &7.19   &1.119 &0.1974 &0.8981 &1.340 &9.80 &242.56 &251.479 &5.55 &7\\[-0.5ex]
               &0.14 &0.009 &0.05    &0.002 &0.0015 &0.0002 &0.004 &0.06 &0.15 & &0.01\\

2018\,Jan\,18 &166.17 &77.396 &16.63  &1.501 &0.4052 &0.89252 &2.109 &25.40 &227.33 &298.365 &4.35 &2.7\\[-0.5ex]
              &0.03 &0.006   &0.02   &0.002 &0.0007 &0.00003 &0.004 &0.02 &0.03 & &0.01\\

2018\,Apr\,08 &252.72 &48.17 &11.89  &0.9888 &0.0962 &0.8936 &1.0839 &22.23 &283.1 &18.634 &6.07 &71\\[-0.5ex]
              &0.03 &0.02      &0.03 &0.0007 &0.0002 &0.0005 &0.0010 &0.05 &0.5 & &0.01\\

2018\,May\,23 &43.35 &78.91 &6.61  &1.112 &0.1501 &0.9453 &1.279 &10.52 &119.4 &62.391 &5.58 &9.5\\[-0.5ex]
              &0.21 &0.09 &0.05  &0.002 &0.0014 &0.0002 &0.004 &0.08    &0.4 & &0.01\\

2018\,Sep\,11 &334.594 &18.20 &20.87 &2.639 &0.7344 &0.7007 &4.58 &16.68 &253.60 &168.851 &2.90 &7\\[-0.5ex]
              &0.008 &0.02 &0.04      &0.015 &0.0016 &0.0003 &0.03 &0.03 &0.02 & &0.01\\

2018\,Oct\,08 &347.694 &11.27 &8.35   &1.410 &0.3681 &0.8909 &1.929 &3.82 &233.31 &195.216 &4.66 &3.5\\[-0.5ex]
              &0.013 &0.04      &0.03 &0.003 &0.0014 &0.0003 &0.006 &0.02 &0.01 & &0.01\\

2018\,Nov\,29 &356.91 &81.618 &23.33 &2.429 &0.6301 &0.89854 &3.959 &35.56 &219.70 &246.617 &3.01 &8\\[-0.5ex]
             &0.05 &0.012 &0.03      &0.008 &0.0012 &0.00006 &0.016 &0.04 &0.02 & &0.01\\
\hline
\end{tabular}
\vspace{-1em}
\end{flushleft}
\end{table*}

\begin{figure*}
\centering
\includegraphics[width=16cm]{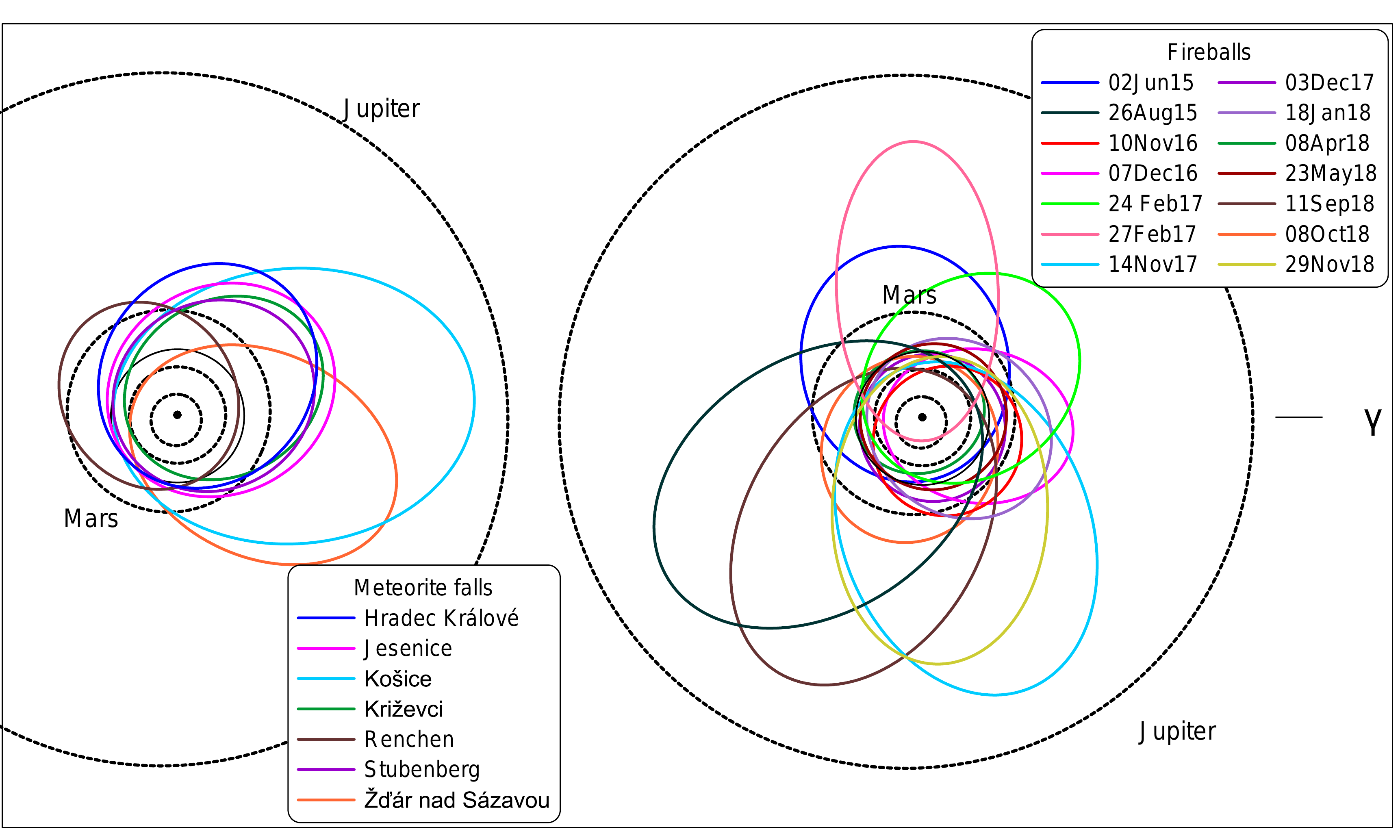}
\caption{Orbits of the meteorites (left) and the fireballs with unrecovered meteorites (right) in the projection to ecliptic plane.
Orbits of planets, except the Earth, are dotted. Vernal equinox is to the right.}
\label{orbitfig}
\end{figure*}

\newpage
\subsection{Orbits}

To complete the information, heliocentric orbits of 14 studied fireballs are given in Table~\ref{orbitlist}. They were computed
by the analytical method of \citet{Ceplecha87} and are compared graphically
with the orbits of the recovered meteorites in Fig.~\ref{orbitfig}. Orbital elements of the meteorites can be found in the original
publications. Their perihelia lie in the Venus-Earth region and the aphelia are located between the orbits of Mars and Jupiter, in most cases
closer to Mars. Except for Renchen, the inclinations are smaller than 10\degr.

The orbits of fireballs have generally the same character. In three cases, the aphelia are within the orbit of Mars, one
of these orbits is of Athen type. On the opposite side, three fireballs had semimajor axes around 2.7~AU and aphelia
around 4.5~AU. This type of orbit is similar to that of the Ko\v{s}ice meteorite. One fireball had low perihelion distance of 0.35~AU 
and two had inclinations above 30\degr. Low perihelion or high inclination leads to higher encounter velocity with the
Earth. Meteoroids in these orbits encounter the same dynamic pressure at higher altitudes than meteoroids in orbits with low
eccentricities and low inclinations and are therefore somewhat disqualified for meteorite survival.

The Tisserand parameters relatively to Jupiter are for some fireballs close or even slightly below the border value of 3.
Nevertheless, there is an overlap of comets and asteroids in this border region \citep{Tancredi} and since all aphelia 
are well within the orbit of Jupiter, we can classify all orbits as asteroidal. 

\begin{table*}
\caption{Computed strewn field coordinates.}
\label{strewndata}
\begin{flushleft}
\vspace{-2em}
\begin{tabular}{lllr@{\hspace{7mm}}lllr@{\hspace{7mm}}lllr}
\hline \hline
Fireball & Longitude & Latitude& Mass & Fireball & Longit. & Latit.& Mass & Fireball & Longit. & Latit.& Mass \\
& \degr E& \degr N & g & & \degr E& \degr N & g & & \degr E& \degr N & g \\
\hline
2015\,Jun\,02 &   11.9976 & 50.1005  &  200  &      2017\,Feb\,27 &  14.4319 & 49.6985& 40        & 2018\,May\,23 &  17.1457 & 49.5317 & 800   \\
&   12.0086 & 50.0890 & 30                   && 14.4143 & 49.6754 & 10               &&17.1533  & 49.5521 & 250                  \\
&  12.0280 &  50.0755 & 3                  & 2017\,Nov\,14 &   ~\,7.0167  & 50.2652 & 11500               && 17.1501   & 49.5733 & 60             \\
2015\,Aug\,26 &   12.4681 & 49.0922 & 125             && ~\,7.7850 &  50.2040 & 5                     && 17.1485 & 49.5856 & 20                   \\
&   12.4517 & 49.0600 & 25                       & 2017\,Dec\,17 & 14.4315  & 48.9888 & 1350          && 17.1471  & 49.5952 & 10                  \\
&  12.4445 & 49.0318 & 4                 && 14.4457 & 48.9957 & 400                              & 2018\,Sep\,11 & 15.6217 & 47.3554 & 40         \\
2016\,Nov\,10 &   21.0824  & 48.2894 & 1100          && 14.4969 & 49.0122 & 10                        && 15.6332 & 47.3145 & 2                     \\
&   21.0778 &  48.3153 & 150                    & 2018\,Jan\,18 & 14.5027 & 49.3063 & 100             & 2018\,Oct\,08 &   14.1967 & 50.3590& 500        \\
&   21.0798 & 48.3338 & 40              && 14.5421 & 49.3165 & 20                                && 14.2330 & 50.3128& 8                     \\
2016\,Dec\,07 & 15.2182 & 49.7128 & 130             && 14.6433 & 49.3430 & 5                        & 2018\,Nov\,29 &   16.6016 & 45.8450 & 900         \\
&   15.1545 & 49.6997 & 10                     & 2018\,Apr\,08 & 16.5482   &46.0773 & 1800            & &16.5991 & 45.8873 & 30                   \\
&   15.1007 &  49.6843 & 2               && 16.5845 & 46.1058 & 600                                                                         \\
2017\,Feb\,24 & 13.2668  & 48.5415 & 40              && 16.5917 & 46.1116 & 500                                                                   \\
&   13.4115 & 48.5154 & 2                      &&16.6400 & 46.1483 & 120                                                                       \\
  &&&     && 16.7213 & 46.2060 & 10                                                                             \\
\hline
\end{tabular}
\vspace{-1em}
\end{flushleft}
\end{table*}

\subsection{Strewn fields}

The computed coordinates of the strewn fields of the 14 studied fireballs are given in Table~\ref{strewndata}. They are provided for the 
case that a meteorite is recovered in the future. The coordinates will enable the meteorite association with one of these fireballs.

For each fireball, the coordinates and approximate mass of the largest expected meteorite are given first. The coordinates were determined
by dark flight computation using the method of \citet{Ceplecha87}. The starting values were based on the observed trajectory
and dynamics at the end of the fireball, fitted by the semi-empirical model. Atmospheric winds were taken from the ALADIN numerical weather
model forecast for the nearest hour and the nearest grid point to the fireball end, kindly provided by R. Bro\v{z}kov\'a from the Czech Hydrometeorological
Institute. Only for the 2018\,Jan\,18 fireball, the radiosonde measurements from Prague (12 UT) were taken (this fall was unfavorable for searches because of low
meteorite mass and very strong gusty winds, which make the landing point predictions uncertain in any case). 

The strewn field is further described by the computed positions of representative smaller fragments, whose existence was inferred from
the fragmentation modeling, primarily from light curve fitting. In cases of fireballs 2017\,Dec\,17, 2018\,Apr\,08, and 2018\,May\,23, the positions
of fragments directly seen in videos are included (2, 3, and 4 fragments, respectively, in addition to the main piece).
In practice, the meteorites can be spread along the central line defined by the listed coordinates but also several kilometers to the sides,
as the experience with well described strewn fields with numerous meteorites shows \citep[e.g.][]{Gnos}.

\section{Discussion}
\label{discussion}

We have shown that fragmentation events during falls of ordinary chondrites do not occur randomly but in two
distinct phases. The first phase occurs at early stages of the atmospheric flight, typically under dynamic pressures of 0.04 -- 0.12 MPa.
In some cases the fragmentation started already at about 0.01 MPa. The first phase was detected in all studied fireballs.
In about 2/3 cases it was catastrophic or nearly catastrophic, i.e.\ more than 40\% of mass was separated from the main body. The first phase was followed
by a quiet period with no or only minor fragmentations. The second phase started at dynamic pressures between 0.9 -- 5 MPa; 
in some cases, especially in smaller meteoroids, already at about 0.5 MPa or even earlier. On the other hand, 
the second phase was sometimes not observed at all. In these cases, however, the meteoroids were decelerated before the
dynamic pressure reached 5 MPa.

\begin{figure}
\centering
\includegraphics[width=9cm]{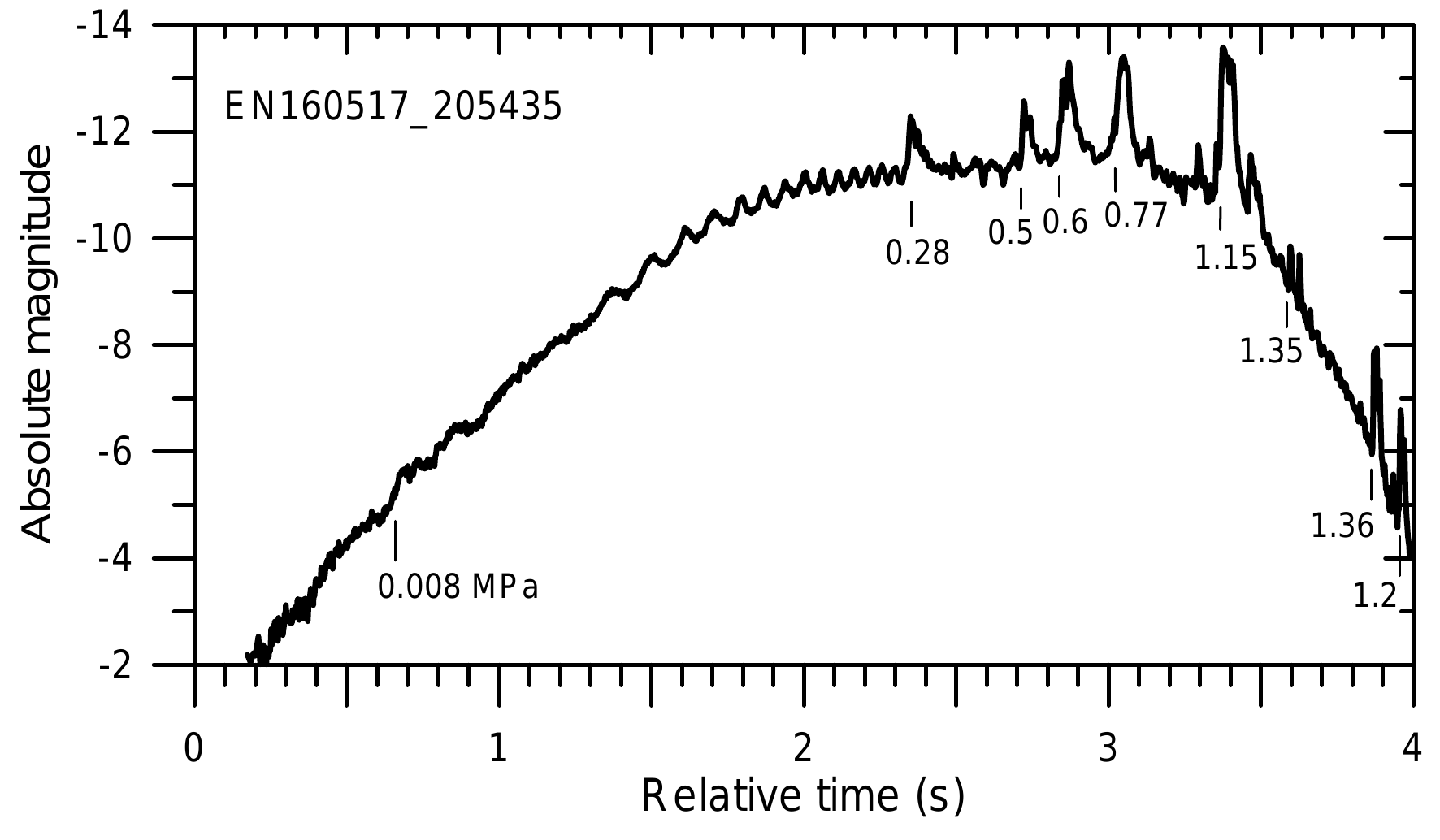}
\caption{Radiometric light curve  of type II fireball EN160517\_205435. Points of main fragmentations and the
corresponding dynamic pressures in MPa are indicated.}
\label{lcII}
\end{figure}

\subsection{Comparison with carbonaceous chondrites}

We emphasize that carbonaceous chondrites behave differently. The recently modeled CM2 meteorite fall Maribo (17 Jan 2009) exhibited
numerous fragmentations along most of the trajectory \citep{Maribo}. The first significant mass loss occurred at 0.017 MPa and 
further disruptions accompanied by large amplitude flares followed at 0.25 -- 4.3 MPa, i.e.\ at a wide range of pressures. 
We identified one similar case among bright fireballs observed by the European Network (though much fainter than Maribo).
The light curve of that fireball, EN160517\_205435, is shown in Fig.~\ref{lcII}. There were numerous flares in the second half 
of the trajectory at dynamic pressures 0.28 -- 1.37 MPa. The shape of the light curve suggests that a minor early fragmentation occurred
as well, possibly at 0.008 MPa. The behavior was similar to Maribo and rather distinct from other fireballs in this study
in the sense that there were numerous well-defined flares covering the whole second half of the light curve.
According to the
PE criterion of \citet{CepMac76}, the fireball was classified as type II. It is therefore probable that it was a carbonaceous chondrite. 
The multi-flare behavior may be characteristic for carbonaceous chondrites, although more data are obviously needed since
Maribo is the only confirmed and well-studied carbonaceous fall.
The entry mass of EN160517\_205435 from the model was 19 kg and the modelling showed that no meteorites larger than few grams can be expected. 
The entry speed was 20.61 km s$^{-1}$ and the orbit was asteroidal.

\subsection{Cracks as the cause of the second phase of fragmentation}

The fragments of ordinary chondrites subject to the second phase of fragmentation have strengths  0.9 -- 5 MPa, 
which is about an order of magnitude lower than the tensile strength of 
ordinary chondritic meteorites \citep[20 -- 40 MPa;][]{Slyuta,Flynn, Ostrowski}. Cracks resulting
from collisions between asteroids (or meteoroids) in interplanetary space can naturally explain such decrease of strength 
in comparison with (nearly) pristine rock. Note that the analysis of Maribo showed that cracks do not play so important role
in carbonaceous chondrites. The reason why carbonaceous chondrites fragment under wide range of pressures is probably that
carbonaceous meteoroids are highly inhomogeneous and contain parts with different degrees of compactness and different 
strengths \citep{Maribo}. Different types of asteroidal material, which have no meteorite analog, also exist. An example was 
the Romanian superbolide of 7 January 2015,
which remained almost intact until 1 MPa and then was quickly pulverized \citep{Romanian}.

The important question is what the reason of the first phase of fragmentation is. Since there is a huge gap (often about an order of magnitude)
in strength between the two phases, it is very unlikely that cracks are responsible for the first phase as well. 
Naturally, cracks of different widths and 3-D shapes can be expected to exist and to lead to different fragmentation strengths.
However, there should be a smooth distribution of crack strengths. The distribution probably covers the second phase, 
i.e.\ 0.9 -- 5 MPa, in some cases, especially in small meteoroids, extending down to 0.5 MPa.
Note that according to widely used \citet{Weibull} distribution, the strength of terrestrial rock decreases with size.
Meteoroids seem to behave differently. We speculate that
the reason of lower strength of smaller meteoroids 
may be that wide cracks producing low strength have higher probability in small meteoroids
to go across the whole body. 

\subsection{The first phase of fragmentation}

The first phase needs to be caused by a different mechanism. A clue may be the Bene\v{s}ov superbolide, which
dropped individual meteorites of at least two different types, LL3.5 and H5 \citep{Benesov}. 
The meteoroid was therefore an assembly of debris of at least two different
asteroids. In this respect it was similar to the Almahata Sitta meteorite fall, which contained even more
meteorite types \citep{Bischoff_AS, Shaddad}. The Bene\v{s}ov meteorite fall was observed by the European Network in 1991. 
Unfortunately, there were
no radiometers at that time and the photographic light curve is not detailed and precise enough for reliable modeling by the semi-empirical model.
From the deceleration at high altitudes, it is, nevertheless, clear that the meter-sized meteoroid disrupted into smaller pieces early
in the flight. The most likely scenario was discussed by \citet{IAU318} and has been depicted in Fig.~\ref{p-m1} together
with other meteorite falls discussed in this paper. The first Bene\v{s}ov disruption most likely occurred at dynamic pressure of 0.05~MPa, 
i.e.\ within the
range for the first phase fragmentation. This can lead to the hypothesis that the strength of the first phase fragmentation
corresponds to the strength with which foreign pieces are held together.

There are no evidences that other meteoroids studied here were composed of pieces of different composition. Nevertheless, also
chemically and mineralogically homogeneous meteoroids can be composed of pieces which were completely separated during
asteroid collisions and then reassembled again. Most of currently existing asteroids in the size range 200 m -- 10 km are 
considered to be gravitational aggregates, so called rubble piles \citep{Pravec, Hestroffer}.  The strength
is therefore near zero. If intergrain van der Waals forces are considered, the strength can reach 
about 25 Pa \citep{Scheeres}. That is still
much lower than observed for the first phase fragmentation. Individual components of meteoroids and 
small asteroids such as Bene\v{s}ov and Almahata Sitta (= asteroid 2008 TC$_3$) must be therefore somehow cemented together
so that the resulting strength is of the order of 50 -- 100 kPa.

\subsection{Formation of low strength meteoroids}

We can only speculate how the cementing can look like. The gravitational reassembly is not supposed to occur
in free space since the mutual gravity of meter-sized or smaller bodies is too low. Instead, the debris were probably first accumulated
on the surface of a larger asteroid, which possibly became subsurface after fallout of more material. Partial impact melting of the
debris and of the adjacent dust grains could cement the material together. The pressure from upper layers could further help.
Note that cementing we are speaking about is relatively weak, only about two times stronger than the tensile strength of snow
\citep[which was measured to be $\sim 0.01$ -- 0.04 MPa by][]{snow}.

The formed breccia resided on the asteroid until further asteroid collision, which finally ejected decimeter and meter-sized fragments into 
free space. Such a multifragmentation history was discussed \citep{Horstmann,Goodrich15} for Almahata Sitta, 
which, nevertheless, is not an ordinary chondrite but predominantly a ureilite. \citet{Goodrich15} proposed that
Almahata Sitta (2008 TC$_3$) originated from the outer layers of regolith of its immediate parent body and other polymict
ureilites, which did not disintegrate into individual mineralogically distinct components during the atmospheric flight, came from deeper
regolith of the same body.  
\citet{Goodrich19} identified two friable and mineralogically diverse Almahata Sitta meteorites probably representing the bulk material
of 2008 TC$_3$, a material that was mostly lost during the atmospheric flight and encompassed the harder material, which more easily
survived as meteorites.

Our study indicates that a history of multiple fragmentation and reaccumulation in interplanetary space
is typical also for ordinary chondrites, even if foreign material is not present and all reaccumulated material originated from the same source.
As a result of this process, the bulk strength of decimeter and meter sized meteoroids was found to be in the range of 0.04 - 0.12 MPa or even less.
It depends on the actual size of the building blocks whether the initial disruption is catastrophic or only small part of mass is lost.
The same properties can be expected for larger bodies of the sizes of tens of meters. Nevertheless, no evidence was found for
fragmentation in the first half of the trajectory of the ordinary chondritic 19-meter Chelyabinsk impactor, except some release
of dust forming atmospheric dust trail \citep{IAU318}. However, the data, especially light curve, are of lower quality, not comparable with
radiometric curves. Moreover, detection of early atmospheric fragmentation of bodies of this size may be difficult.
\citet{Shuvalov} demonstrated that a strengthless body will produce almost identical light curve in the upper part of the trajectory as
the real object.

\section{Summary}
\label{summary}

In this paper we presented the semi-empirical model of atmospheric fragmentation of meteoroids
and its application to 7 confirmed meteorite falls and 14 fireballs with predicted but unrecovered meteorites. 
All recovered meteorites were ordinary chondrites. From the unrecovered meteorites, at least 13 are expected 
on the basis of fall statistics to be ordinary chondrites. We provided additional information, 
orbits and strewn fields, for the fireballs. Details about the meteorite falls can be found in the original papers.

The fact that the strength of meteoroids is significantly lower than the strength of meteorites measured in the laboratory
was emphasized already by \citet{Popova}. The new results of this study is that meteoroid strengths are not distributed randomly
but are predominantly concentrated in two regions (marked B and C below). 
When combined with the data on meteorites and medium-sized asteroids,
we conclude that ordinary chondritic material and bodies can be found in four strength categories. Their typical tensile/fragmentation
strength and probable physical structure is as follows:

\begin{description} 

\item[A (20 -- 40 MPa)] Pristine material or compact breccias formed in asteroids at high pressures. Encountered
as meteorites, i.e.\ the strongest parts of meteoroids, which survived the atmospheric passage. Only rarely
the whole meteoroid can be of this type (the example is Carancas). Microscopic cracks can be present.

\item[B (0.5 -- 5 MPa)] Cracked material. Macroscopic cracks were formed during asteroid collisions in interplanetary
space. The body fragments along the cracks into category A material during the atmospheric flight when dynamic pressure reaches the
strength value. In favorable conditions (low entry speed and shallow angle), dynamic pressure may not reach the
strength value and the body can land as a meteorite.

\item[C (0.04 -- 0.12 MPa)] Reassembled and cemented material. According to our hypothesis, which needs to be tested,
this strength category corresponds to material which was secondarily formed from debris of asteroid collisions
(of categories A and B) cemented together on the surface or near-surface layers of asteroids. During subsequent collisional evolution, 
the layers were destroyed and its parts were released into interplanetary space as individual meteoroids.
During the atmospheric entry, they are early separated into the A and B category components.

\item[D ($\sim 0$ Pa)] Reassembled material held together only by mutual gravity or van der Waals forces. This category
corresponds to rubble pile asteroids. Meteoroids of type D have not been observed.

\end{description}

The fact that most of meteoroids are weakly cemented aggregates of cracked material is the reason why the terrestrial
atmosphere protects us effectively against meteoroid impacts. For example, a 1000 kg meteoroid of category A entering
at 15 km s$^{-1}$ with entry angle 45\degr\ would produce a 600 kg meteorite (assuming the nominal
ablation coefficient of 0.005 kg MJ$^{-1}$). But the reality is different. In a typical case, hundreds of mostly 
small meteorites of total mass less than 100 kg are produced instead.
\vspace{-3mm}

\acknowledgements

The authors acknowledge the work of all operators and technicians who
are keeping the observatories of the European Fireball Network and the associated infrastructure running. 
Special thanks go to J. Svore\v{n} (Astronomical Institute of the Slovak 
Academy of Sciences) and the late H. Mucke (\"Osterreichische Astronomische Verein) for their support of the network 
in Slovakia and Austria, respectively. We are indebted to R. Bro\v{z}kov\'a (Czech Hydrometeorological
Institute, CHMI) for providing us meteorological models needed for dark flight computation.
We thank D. \v{S}\v{c}erba, L. Ronge  (CHMI), D. \v{S}egon (Croatian Meteor Network), W. Stelzig, and M. Landy Gyebnar
for providing their fireball images or videos for analysis.
We appreciate the tireless meteorite searching by R. Sporn and M. Neuhofer and the coordination work of D. Heinlein.
This work was supported by the grant 19-26232X from the Czech Science Foundation (GA\,\v{C}R).
The recent modernization of the network, crucial for this work, was funded from the Praemium Academiae of the Czech Academy
of Sciences. The institutional research plan is RVO:67985815.

\end{document}